\documentclass[lettersize,journal,twoside]{IEEEtran}
\usepackage{amsmath,amsfonts}
\usepackage{array}
\usepackage[caption=false]{subfig}
\usepackage{textcomp}
\usepackage{stfloats}
\usepackage{url}
\usepackage{verbatim}
\usepackage{graphicx}
\usepackage{cite}
\usepackage{siunitx}
\usepackage{mathrsfs}
\usepackage{mathtools}
\mathtoolsset{showonlyrefs=true}
\usepackage{bm}
\usepackage{fancybox}
\usepackage{algpseudocode}
\usepackage{multirow}

\begin{document}

\title{Cyber-Secure Teleoperation With Encrypted Four-Channel Bilateral Control}

\author{Haruki Takanashi,
Akane Kosugi, 
Kaoru Teranishi,
Toru Mizuya,
Kenichi Abe, and
Kiminao Kogiso
\thanks{This work was supported in part by JSPS KAKENHI 22H01509.
This paper will be presented in part at the IEEE/SICE International Symposium on System Integrations (SII 2023) held in Atlanta, GA, USA, January 17-20, 2023. \textit{(Corresponding author: Haruki Takanashi.)}}
\thanks{Haruki Takanashi, Akane Kosugi, and Kiminao Kogiso are with the Department of Mechanical and Intelligent Systems Engineering, The University of Electro-Communications, Chofu, Tokyo 1828585, Japan (e-mail: takanashi@uec.ac.jp; kosugi@uec.ac.jp; kogiso@uec.ac.jp).}
\thanks{Kaoru Teranishi is with the Department of Mechanical and Intelligent Systems Engineering, The University of Electro-Communications, Chofu, Tokyo 1828585, Japan, and is also a Research Fellow of Japan Society for the Promotion of Science, Chiyoda-ku, Tokyo 1020083, Japan (e-mail: teranishi@uec.ac.jp).}
\thanks{Toru Mizuya and Kenichi Abe are with Kanagawa Institute of Industrial Science and Technology, Ebina, Kanagawa 2430435, Japan (e-mail: tmizuya@kistec.jp; ken1abe@kistec.jp).}
}

\maketitle

\begin{abstract}
  This study developed an encrypted four-channel bilateral control system that enables posture synchronization and force feedback for leader and follower robot arms.
  The encrypted bilateral control system communicates encrypted signals and operates with encrypted control parameters using homomorphic encryption.
  We created two-axis robot arms and identified them to obtain a nonlinear model consisting of a linear system and a nonlinear disturbance.
  Disturbance and reaction-force observers and a proportional-derivative controller were designed to construct a networked robot-manipulation system.
  The controllers in the constructed bilateral control system were securely implemented on dedicated computers using a controller encryption technique.
  The experimental results demonstrate that the developed bilateral control system can facilitate encrypted communication and controller, synchronize the posture, and feed the reaction force back.
  Through experimental validation, the encrypted four-channel bilateral control system enables the inheritance of control performance from the original (unencrypted) bilateral control system.
\end{abstract}

\begin{IEEEkeywords}
  Encrypted control, experimental validation, four-channel bilateral control, robot arm, teleoperation.
\end{IEEEkeywords}

\section{Introduction}
\IEEEPARstart{B}{ILATERAL} control is an important technology for the handling of hazardous materials~\cite{Clement1985}, space and deep-sea exploration~\cite{Bejczy1987, Lei2011, Funda1991}, and telesurgery~\cite{Rovetta1996, Jia2015}.
The bilateral control system consists of a human-operated leader and a follower that follows the leader’s posture to present the remote environment of the follower to the leader.
Various architectures have been proposed to realize bilateral control, such as~\cite{HOKAYEM2006} and its references.
Well-known architectures include the position-position~\cite{Niemeyer1991, Tachi1991}, force-position~\cite{Hannaford1988, Tachi1991, Willaert2009}, and four-channel~\cite{Lawrence1993, Yokokhji1994, Matsumoto:2003aa, Iida2004, Natori2006, Yang2020, Hangai2021} controllers.
The position-position controller aims to synchronize the positions of the leader and follower, 
the force-position controller presents the reaction force of the follower to the leader, and the four-channel controller communicates both positions and forces for the leader and follower.
Recent developments in communication technology have enabled teleoperation over the Internet, and subsequently, the problems of communication delays, packet losses, and cybersecurity must be overcome.
Numerous studies on communication delays and packet loss in teleoperation have been conducted~\cite{Funda1991, Natori2004, Ueda2004, HOKAYEM2006, Varkonyi2014, islam2015, Beerens2020}, whereas cybersecurity issues have recently emerged because of increased cyber-attacks against industrial control systems, such as Stuxnet and Industroyer~\cite{Stuxnet2011, Industroyer2016}.

% Motivation
As the threat of cyber-attacks increases, preparing and enhancing the cyber security of bilateral control systems is essential.
Several studies have been conducted on the modeling and classification of cyber-attack methods, such as denial-of-service (DoS) attacks and packet drop attacks on force-position-type bilateral control systems~\cite{Munteanu2018}, specific cyber-attacks on surgical robots and remote rescue robots~\cite{Asif2020, Bonaci2015}, and false data injection attacks on practical force-position-type bilateral control systems~\cite{Dong2020, Munteanu2018}.
Moreover, several studies have been conducted on cyber-attack countermeasures for bilateral control systems, which are broadly classified into proactive and reactive countermeasures.
Proactive countermeasures include secure communication protocols, such as a protocol that considers both security and quality of service~\cite{kalam2016}, an adaptive information coding scheme to support confidentiality and adaptive reliability simultaneously~\cite{Tozal2011}, and an adaptive controller to achieve globally asymptotic stability of the force-position-type bilateral control systems under DoS attacks~\cite{Jiang2020}.
For the reactive countermeasures, the authors of~\cite{Dong2020} proposed a physics-based attack detection scheme with an encoding–decoding structure for general false data injection attacks on force-position-type bilateral control.
Another promising technology that achieves both proactive and reactive countermeasures is the integration of cryptography into control systems, called encrypted control~\cite{Kogiso2015, FAROKHI2016163, Kim2016175, schlze2021en, Andreea2017}.
Encrypted control enables the concealment of communication data and control parameters, and makes it easier to detect falsification and injection attacks.
For example, \cite{shono2022} applied encrypted control to an experimental pneumatically actuated force-position-type bilateral control system.
The authors of~\cite{KOSIERADZKI2022se} demonstrated the feasibility of encrypting the entire motion-control scheme of a teleoperated system in a simulation.
\IEEEpubidadjcol

However, although all these studies are for position-position or force-position-type bilateral control, they are inappropriate for four-channel bilateral control.
The four-channel type achieves transparency, that is, the presentation of a sense of the remote environment to the user~\cite{Lawrence1993, Yokokhji1994} and can scale the position and force between the leader and follower~\cite{Zhu1995}.
It is essential to develop a four-channel bilateral control system that considers cybersecurity.

\begin{figure}[!t]
\centering
\includegraphics[width=1.0\linewidth]{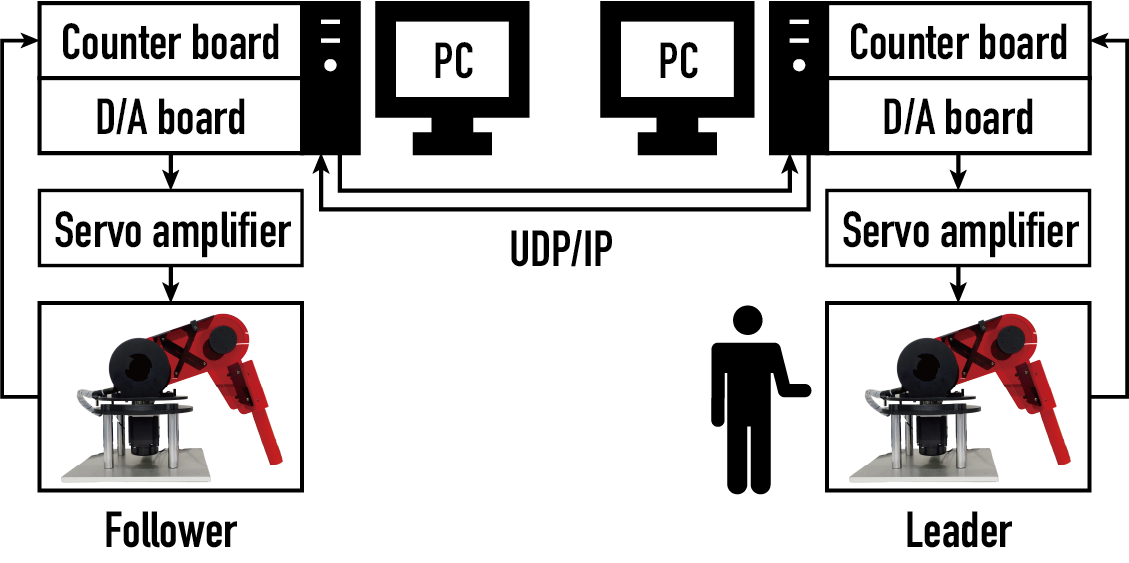}
\caption{Configuration diagram of proposed networked robot manipulation system. \label{fig:concept}}
\end{figure}
\begin{figure}[!t]
\centering
\includegraphics[width=1.0\linewidth]{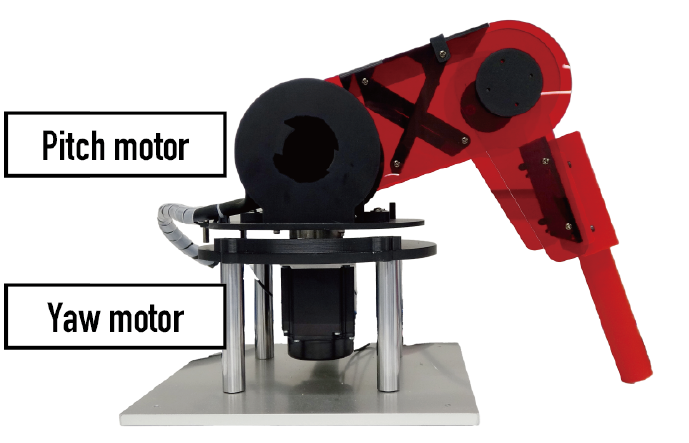}
\caption{Side view of the two-axis robot arm.}
\label{fig:robotarm}
\end{figure}

The purpose of this study is to develop an encrypted four-channel bilateral control system that enables synchronization of position and force feedback using robot arms.
The bilateral control system developed in this study, shown in Fig.~\ref{fig:concept}, configures the leader and follower two-axis robot arms, shown in Fig.~\ref{fig:robotarm}, and the operator manipulates the leader robot arm.
The robotic arms were controlled using a computer.
The bilateral control system was constructed to achieve the following three objectives:
1) the posture of the leader and follower are synchronized, 
2) the reaction force on the follower is fed back to the leader,
and 3) the signals on the communication links and parameters inside the controllers are encrypted with a partially (multiplicative) homomorphic encryption.
To realize the encrypted four-channel bilateral control system, shown in Fig.~\ref{fig:concept}, the unencrypted four-channel bilateral control system proposed in \cite{Matsumoto:2003aa} was constructed using dedicated two-axis robot arms.
The model and parameters of the robot arms were identified and applied to a model-based bilateral control system design.
A reaction-force observer was used to estimate the reaction force without a force sensor.
Next, the leader and follower controllers of the four-channel bilateral control system were described in state-space representation and securely implemented using the controller encryption techniques proposed in~\cite{Kogiso2015,Teranishi2020en}.
The resulting control system is the first application of four-channel bilateral control in the encrypted control fashion.
Moreover, we verified the three requirements of the encrypted bilateral control system by conducting two experiments, namely motion tracking for posture synchronization and force feedback when the follower contacts obstacles such as a sponge and an aluminum prism for an operator to feel the difference in hardness.
Moreover, the effects of encryption on the computation time and quantization error were investigated to confirm that the computation time was within the sampling period and the quantization error was negligibly small.
Through experimental verification, this study confirmed that an encrypted four-channel bilateral control system can be developed.

The contributions of this study are as follows:
\begin{enumerate}
    \item We developed an encrypted four-channel bilateral control system with two-axis robot arms by conducting a secure implementation of the leader and follower controllers. 
    The controller encryption technique enables the communications and controller parameters to be kept hidden, which helps establish a cybersecurity-enhanced bilateral control system.
    \item The developed bilateral control system was experimentally verified to maintain the tracking and force feedback performance before and after encryption.
    The secure implementation enables us to inherit the control performance from that of the original (unencrypted) bilateral control system.
\end{enumerate}

The remainder of this paper is organized as follows.
Section \ref{sec:robot} describes the model of the robot arm, configuration of the networked robot manipulation system, and problem setting of this study.
Section \ref{sec:four-channel} introduces the state-space representation of a four-channel bilateral control system. 
Section \ref{sec:encryption} presents the encryption of the four-channel bilateral controllers.
Section \ref{sec:experiment} demonstrates that the developed bilateral control system can be used to conduct the robot teleoperation.
Section \ref{sec:conclusion} summarizes the study and describes future works.

%%%
\section{robot arm and its model\label{sec:robot}}
This section describes the teleoperation system and robot arm model developed in this study.
Section~\ref{sec:networkrobot} describes the configuration and specifications of the teleoperation system, and Section~\ref{sec:modeling} presents a mathematical model of the robot arm and its disturbance.

%%%
\subsection{Networked robot arms~\label{sec:networkrobot}}
In this study, we developed a teleoperation system using two robotic arms, the leader and follower arms, as shown in Fig.~\ref{fig:concept}.
A human operator operates the leader.
The leader and follower were each connected to a computer.
The leader and follower computers were connected via a wired LAN and communicated via UDP/IP socket communication.

The specifications for the robots and computers are listed in TABLE~\ref{tab:spec}. 
In this study, we used the robotic arm shown in Fig.~\ref{fig:robotarm}.
The robot arm is driven by two AC servo motors, a yaw axis motor, and a pitch axis motor, and the input and output of each motor are the current and angle, respectively.
The control input calculated by the computer was transmitted to the motor of the robot arm through the D/A board and servo amplifier.
The angle of each motor was read by the rotary encoder attached to the motor and transmitted to the computer via a servo amplifier and counter board.
CentOS 8.3, compatible with RedHat Enterprise Linux (RHEL), was used as the operating system of the computer. 
In addition, the Advanced Robot Control System V6 (ARCS6)~\cite{ARCS} was used as the framework for real-time computation. 

%%%
\subsection{Modeling~\label{sec:modeling}}
The yaw and pitch axes of the robot arm were orthogonal; thus, the equations of motion for each motor can be independently established.
The equations of motion for each motor are as follows:
\begin{equation}
\bar{J}_{j}\ddot{\theta}_j(t) = \bar{K}_ji_j(t) - \tau_j^{d}(t), \quad \forall j\in\{1,2\}, \label{math:yaw} 
\end{equation}
\hspace*{-1ex}
where the subscript $j\in\{1, 2\}$ denotes the yaw and pitch motors, respectively, 
$i_j\in\mathbb{R}$ is the current input to the motor, 
$\theta_j\in\mathbb{R}$ is the angle measurable by the encoder, and $\tau_j^{d}\in\mathbb{R}$ is the unknown disturbance torque.
$\bar{J}_{j}$ and $\bar{K}_j$ are the nominal values of the moment of inertia and torque coefficient, respectively, and their values are listed in TABLE~\ref{tab:spec}, which are obtained from the specifications. 
Hereafter, subscript $j$ is omitted for simplicity.

In this study, the torque was assumed to consist of the following components: modeling error, cable tension, gravity, friction, and an external force, that is,
\begin{equation}
\tau^{d}(t) := (J - \bar{J}) \ddot{\theta}(t) + (\bar{K} - K) i(t) + f(\theta, \omega) + \tau^{e}(t),
\label{math2:taud}
\end{equation}
where $J$ is the moment of inertia; $K$ is the torque coefficient of the motor;
$f$ is a nonlinear function of the angle $\theta$ and velocity $w=\dot\theta$ that involves the cable tension, gravity, and friction forces, which need to be identified in the experiment; and 
$\tau^{e}$ is the external force, which is estimated by the reaction force observer~\cite{Murakami:1993aa}.

\begin{table}[tb]
\caption{Control system specifications.}
\label{tab:spec}
\centering
\begin{tabular}{ll}
\hline
\textbf{Yaw axis motor} & MITSUBISHI HK-KT-43W \\
Servo amplifier & MITSUBISHI MR-J5-40A \\
Rated power & \SI{400}{W} \\
Rated torque & \SI{1.3}{Nm} \\
Rated Speed & \SI{3000}{rpm} \\
Rated Current & \SI{2.6}{A} \\
Moment of inertia $\bar{J}_{1}$ & ${0.410\times10^{-4}}$\si{kgm^2} \\
Torque coefficient $\bar{K}_{t1}$ & $\SI{0.5}{Nm/A}$ \\
Pulse per rotation & \SI{131072}{pulses/rev} \\
\hline
\textbf{Pitch axis motor} & MITSUBISHI HK-KT-7M3W \\
Servo amplifier & MITSUBISHI MR-J5-70A \\
Rated power & \SI{750}{W} \\
Rated torque & \SI{2.4}{Nm} \\
Rated Speed & \SI{3000}{rpm} \\
Rated Current & \SI{4.7}{A} \\
Moment of inertia $\bar{J}_{2}$ & ${1.37\times10^{-4}}$\si{kgm^2} \\
Torque coefficient $\bar{K}_{t2}$ & $\SI{0.5106}{Nm/A}$ \\
Pulse per rotation & \SI{131072}{pulses/rev} \\
\hline
\textbf{PC} & \\
CPU & Intel Core i7-10700K (\SI{5.1}{GHz}) \\
Memory & \SI{16}{GB} \\
OS & CentOS 8.3 \\
D/A Board & Interface PEX-340216 (\SI{16}{bit}) \\
Counterboard & Interface PEX-632104 (\SI{32}{bit}) \\
\hline
\end{tabular}
\end{table}

%%%
\section{four-channel bilateral control system\label{sec:four-channel}}
\begin{figure*}
\centering
\includegraphics[keepaspectratio, width=1.0\linewidth]{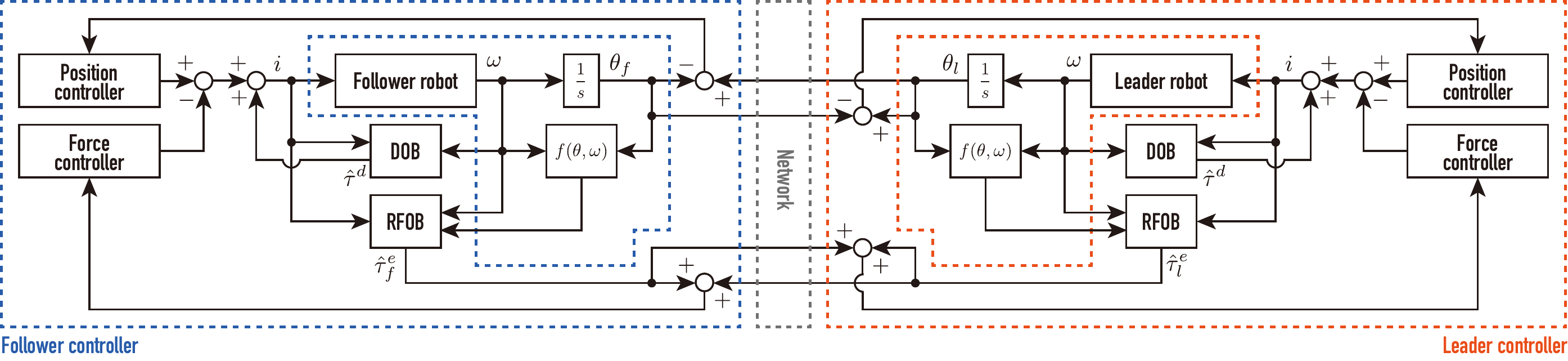}
\caption{Block diagram of the four-channel bilateral control system.}
\label{fig:bd_4ch_a}
\end{figure*}

In this section, we derive the state-space representation of the four-channel bilateral control system shown in Fig.~\ref{fig:bd_4ch_a}. 
The gray dotted line at the center of the figure indicates the network between the leader and follower.
In the network, four signals are transmitted and received: the rotation angle $\theta_l$ and the estimated reaction force $\hat{\tau}^e_l$ of the leader, and the rotation angle $\theta_f$ and the estimated reaction force $\hat{\tau}^e_f$ of the follower.

We derived the disturbance observer (DOB) to estimate the disturbance applied to the robot in Section~\ref{sec:dob} and the reaction force estimation observer to estimate the contact force with the environment in Section~\ref{sec:rfob}. 
Using the DOB, we estimated a model of the nonlinear terms $f(\theta,\omega)$ in Section~\ref{sec:residual} and combine them to derive a state-space representation of the controller of the system in Fig.~\ref{fig:bd_4ch_a}.

\subsection{Disturbance observer (DOB) \label{sec:dob}}
The DOB~\cite{Ohnishi1996mc} compensates for disturbances in the robot arm by estimating and feeding back the disturbance from the input current and velocity.
The DOB cancels the modeling error of each motor, allowing each motor to be nominalized.
Assuming that the disturbance torque is constant, the motor \eqref{math:yaw} is given as follows:
\begin{align}
  \dot{x}(t) &= A x(t) + B i(t), \quad \omega(t) = C x(t), \label{math3:model}
\end{align}
\begin{align}
  A &= \begin{bmatrix}
    0 & -\frac{1}{\bar{J}} \\
    0 & 0
  \end{bmatrix},~
  B = \begin{bmatrix}
    \frac{\bar{K}}{\bar{J}} \\
    0
  \end{bmatrix},~
  C = \begin{bmatrix}
    1 & 0
  \end{bmatrix},~
  x = \begin{bmatrix}
    \omega \\
    \tau^{d}
  \end{bmatrix}.
\end{align}
The rank of the observability matrix of the system, represented by \eqref{math3:model}, is
\begin{align}
  \mathrm{rank} \begin{bmatrix}
    C \\
    CA
  \end{bmatrix} &= \mathrm{rank} \begin{bmatrix}
    1 & 0 \\
    0 & -\frac{1}{\bar{J}}
  \end{bmatrix} = 2.
\end{align}
Thus, the system was observable.

In this section, we use the method developed by Gopinath~\cite{gopinath1971} to construct a DOB for the system  \eqref{math3:model}.
If $D \coloneqq \begin{bmatrix}
  0 & 1
\end{bmatrix}$, the transformation matrix $T$ is given as:
\begin{align}
  T &= \begin{bmatrix}
    C \\
    D
  \end{bmatrix}
  = \begin{bmatrix}
    1 & 0 \\
    0 & 1
  \end{bmatrix}.
\end{align}
Hence, the state variables and parameters in \eqref{math3:model} can be converted into
\begin{align}
  T x &= x = \begin{bmatrix}
    \omega \\
    \tau^{d}
  \end{bmatrix},~ \label{math3:tx}\\
  T A T^{-1} &= A = \begin{bmatrix}
    A_{11} & A_{12} \\
    A_{21} & A_{22}
  \end{bmatrix} = \begin{bmatrix}
    0 & -\frac{1}{\bar{J}} \\
    0 & 0
  \end{bmatrix},~ \label{math3:ta}\\
  T B &= B = \begin{bmatrix}
    B_1 \\
    B_2
  \end{bmatrix} = \begin{bmatrix}
    \frac{\bar{K}}{\bar{J}} \\
    0
  \end{bmatrix},~ \label{math3:tb}\\
  C T^{-1} &= C = \begin{bmatrix}
    1 & 0
  \end{bmatrix}. \label{math3:tc}
\end{align}
Using these matrices, we obtain
\begin{subequations} \label{math3:model34}
\begin{align}
  \dot{\tau}^{d}(t) = A_{22} \tau^{d}(t) + A_{21} \omega(t) + B_2 i(t), \label{math3:model3}\\
  \dot{\omega}(t) - A_{11} \omega(t) - B_1 i = A_{12} \tau^{d}(t). \label{math3:model4}
\end{align}
\end{subequations}
The observer of the systems \eqref{math3:model3} and \eqref{math3:model4} is as follows:
\begin{align}
  \dot{\hat{\tau}}^{d}(t) &= A_{22} \hat{\tau}^{d}(t) + A_{21} \omega(t) + B_2 i(t) \\
  &\quad + L (\dot{\omega}(t) - A_{11} \omega(t) - B_1 i(t) - A_{12} \hat{\tau}^{d}(t)),
\end{align}
where $L$ denotes the observer gain.
If $q(t) \coloneqq \hat{\tau}^{d}(t) - L \omega(t)$ and $L = -\bar{J} g_d$, then
\begin{align}
  \dot{q}(t) &= A_{22} (q(t) + L \omega(t)) + A_{21} \omega + B_2 i(t) \\
  &\quad + L (\dot{\omega}(t) - A_{11} \omega(t) \\
  &\quad - B_1 i(t) - A_{12} (q(t) + L \omega(t))) - L \dot{\omega}(t) \\
  &= (A_{22} - L A_{12}) q(t) + (B_2 - L B_1) i(t) \\
  &\quad + (A_{21} - L A_{11}+ L A_{22} - L^2 A_{12}) \omega(t) \\
  &= -g_d q(t) + \bar{K} g_d i(t) + \bar{J} g_d^2 \omega(t) \label{math3:qdot}\\
  \begin{bmatrix}
    \omega(t) \\
    \hat{\tau}^{d}(t)
  \end{bmatrix} &= T^{-1} \begin{bmatrix}
    0 \\
    1
  \end{bmatrix} q(t) + T^{-1} \begin{bmatrix}
    1 \\ 
    L
  \end{bmatrix} \omega(t) \\
  &= \begin{bmatrix}
    \omega(t) \\
    q(t) - \bar{J} g_d \omega(t)
  \end{bmatrix} \label{math3:tau}
\end{align}
are obtained.
Thus, the DOB of the system \eqref{math3:model} is given as follows:
\begin{subequations}
\begin{align}
  \dot{q}(t) &= -g_d q(t) + \bar{K} g_d i(t)  + \bar{J} g_d^2 \omega(t), \label{math3:dob} \\
  \hat{\tau}^{d}(t) &= q(t) - \bar{J} g_d \omega(t) \label{math3:tauhat}.
\end{align}
\end{subequations}
The DOB block of the leader and follower controllers shown in Fig.~\ref{fig:bd_4ch_a} contains the state-space representation of the disturbance observer in~\eqref{math3:tauhat}.
The DOB block has a state variable $q$, inputs current $i$ and angular velocity $\omega$ and outputs a disturbance estimate $\hat{\tau}^d$.
In this study, the cut-off frequency of the DOB $g_d$ was set to $\SI{100}{rad/s}$.

%%%
\subsection{Reaction force observer\label{sec:rfob}}
The reaction force observer (RFOB) estimates the external force applied to the control target from the input current, velocity, and models of the frictional force and gravity.
Assume that the disturbance on the plant is expressed using \eqref{math2:taud} and is estimated by the DOB. 
In this case, the external force applied to the motor is
\begin{align}
  \tau^{e}(t) &= \tau^{d}(t) - (J - \bar{J}) \ddot{\theta}(t) - (\bar{K} - K) i(t) - f(\theta, \omega). \label{math4:tauext}
\end{align}
Assuming that the difference between the nominal and true values of the moment of inertia and torque coefficient is negligibly small, the following equation holds:
\begin{align}
  \tau^{e}(t) &= \tau^{d}(t) - f(\theta, \omega).
  \label{math4:tauext2}
\end{align}
From the literature~\cite{Murakami:1993aa} and Equations \eqref{math3:dob} and \eqref{math4:tauext2}, 
the RFOB is given as follows:
\begin{align}
  \dot{z}(t) &= -g_r z(t) + K g_r i(t) + \bar{J} g_r^2 \omega(t) - g_r f(\theta, \omega),  \label{math4:rfob}\\
  \hat{\tau}^{e}(t) &= z(t) - \bar{J} g_r \omega(t) \label{math4:tauexthat}.
\end{align}
The RFOB block of the leader and follower controllers shown in Fig.~\ref{fig:bd_4ch_a} contains the state-space representation of the RFOB in ~\eqref{math4:tauexthat}.
The RFOB block has a state variable $z$, inputs the current $i$, angular velocity $\omega$, and nonlinear term $f(\theta, \omega)$, and outputs an estimate of the reaction force $\hat{\tau}^e$.
In this study, the cut-off frequency of the RFOB $g_r$ was set to $\SI{100}{rad/s}$.

%%%
\subsection{Estimation of nonlinear terms\label{sec:residual}}
To determine the external forces applied to the robot arm, nonlinear terms $f(\theta, \omega)$ such as cable tension, gravity, and frictional force, must be determined.
First, a static test was performed to estimate the tensile and gravitational forces acting on the cable. 
In this test, each motor was made to stand still at certain angles and the disturbance torque was recorded.
The sampling period was $\SI{5}{ms}$, yaw axis was moved from $\SI{-120}{deg}$ to $\SI{120}{deg}$, and pitch axis was moved from $\SI{10}{deg}$ to $\SI{80}{deg}$ by $\SI{10}{deg}$. 
The disturbance torque was recorded at each angle by stopping $\SI{10}{s}$.

Subsequently, a constant-velocity test was conducted to estimate the frictional force. 
In this test, each motor was moved at a constant angular velocity and the disturbance torque was recorded.
The sampling period was $\SI{5}{ms}$ and the angular velocity was set from $\SI{0.05}{rad/s}$ to $\SI{0.5}{rad/s}$ by $\SI{0.05}{rad/s}$ and from $\SI{0.5}{rad/s}$ to $\SI{1.0}{rad/s}$ by $\SI{0.1}{rad/s}$.
Ten experiments were conducted for each axis and angular velocity.

\begin{figure}[!t]
\centering
\subfloat[]{\includegraphics[scale=.54]{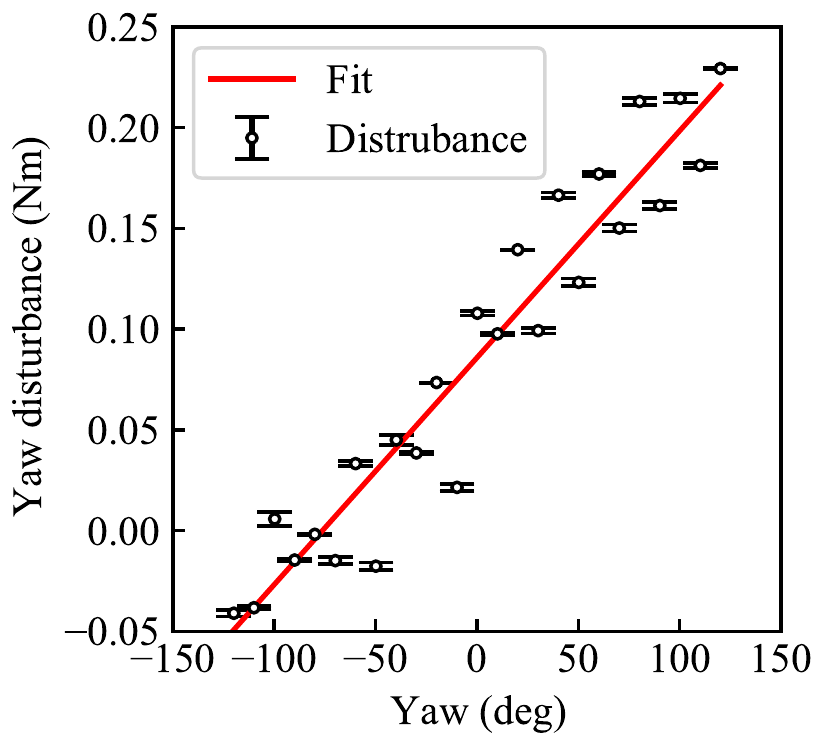}%
\label{fig:yawcable}}
% \hfil
\subfloat[]{\includegraphics[scale=.54]{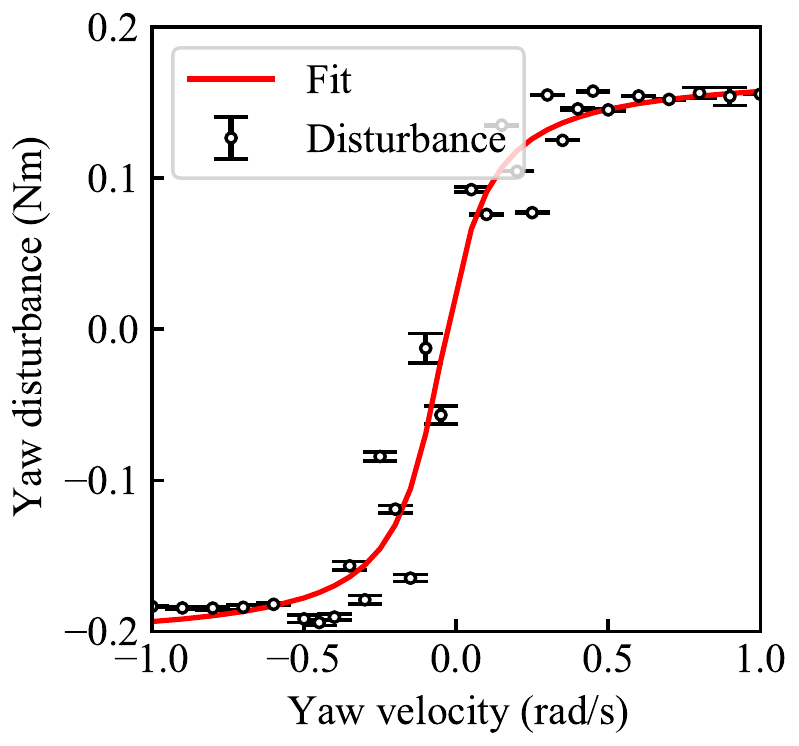}%
\label{fig:yawfric}} \\
\subfloat[]{\includegraphics[scale=.54]{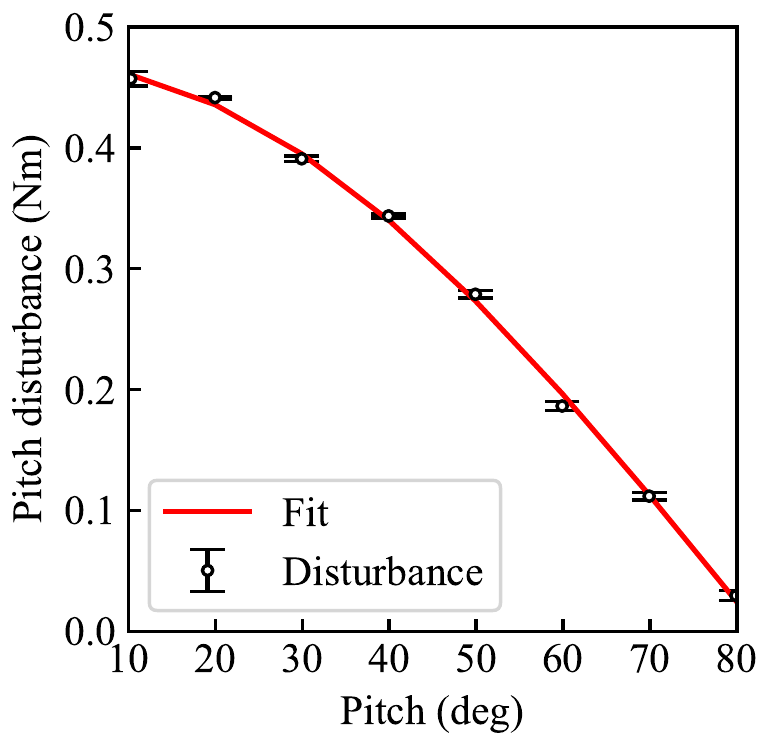}%
\label{fig:pitchg}}
% \hfil
\subfloat[]{\includegraphics[scale=.54]{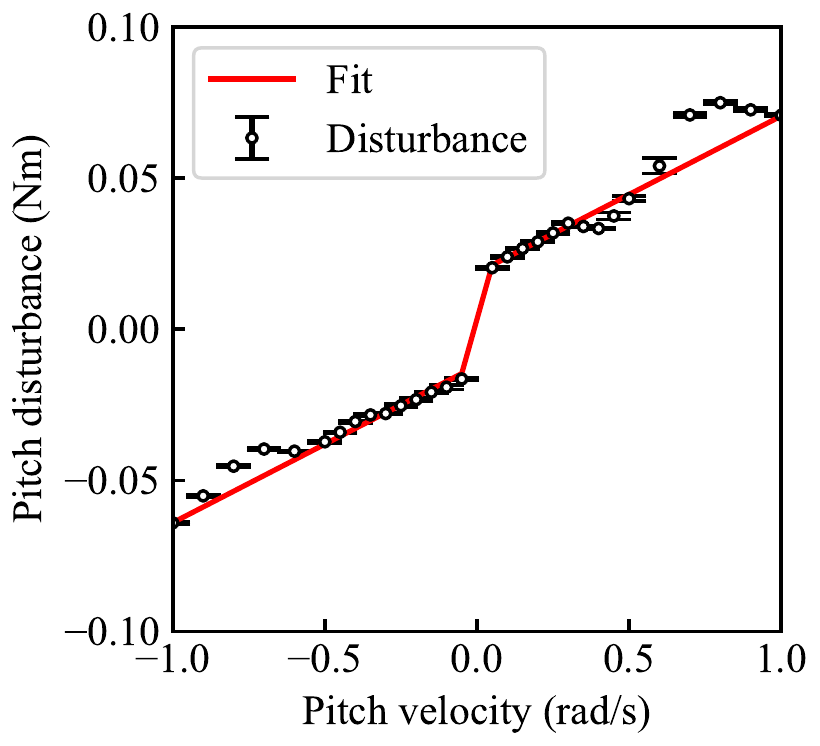}%
\label{fig:pitchfric}}
\caption{Estimated disturbance model. (a) Estimated cable tension model of yaw axis motor. (b) Estimated friction model of yaw axis motor. (c) Estimated gravity model of pitch axis motor. (d) Estimated friction model of pitch axis motor.}
\label{fig:f}
\end{figure}

Figs.~\ref{fig:f} shows the results of the static and constant-velocity tests using the model obtained in the static test.
The black and red lines represent the mean and standard deviation of the estimated disturbance torque, and its approximate curve, respectively.

From the results of the static and constant-velocity tests, the cable tension, gravity, and friction forces on the yaw and pitch axises are expressed as follows:
\begin{align}
	f(\theta_1, \omega_1) &= a_{c} \theta_1+ b_{c} + a_{f1} \tan^{-1}{(b_{f1}\omega_y + c_{f1})} + d_{f1}, \\
	f(\theta_2, \omega_2) &= a_{g} \sin{(b_g \theta_2 + c_g)} + a_{f2} \tan^{-1}{(b_{f2}\omega_2 + c_{f2})} \\
	&\quad + d_{f2} \omega_2 + e_{f2}.
\end{align}
The coefficient values obtained in the test are listed in TABLE~\ref{tab:four-channelparams}.

\subsection{Unencrypted four-channel bilateral control system\label{sec:controller}}
\begin{figure}[!t]
\centering
\includegraphics[width=1.0\linewidth]{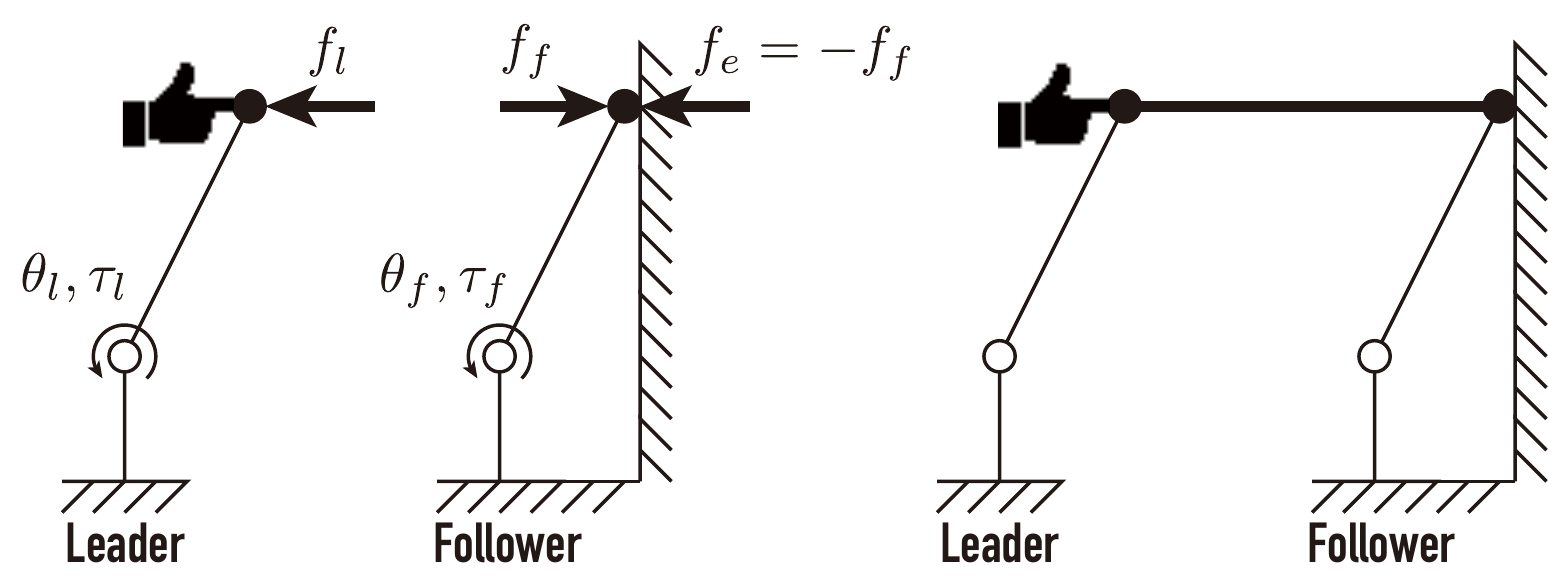}
\caption{Conceptual diagrams of the bilateral controls.}
\label{fig:bilateralctrl}
\end{figure}

This section derives the state-space representation of the four-channel bilateral control with a DOB and RFOB.
Fig.~\ref{fig:bilateralctrl} shows a conceptual diagram of four-channel bilateral control.
Note that the subscripts $l$, $f$, and $e$ denote the leader, follower, and environment, respectively.
The control object for four-channel bilateral control is denoted as follows:
\begin{subequations}
\begin{align}
\theta_l - \theta_f &= 0 \label{math5:theta}\\
\tau^{e}_l + \tau^{e}_f &= 0 \label{math5:tau}
\end{align}
\end{subequations}
where $\tau^{e}_l$ is the torque applied by the human operator to the leader and $\tau^{e}_f$ is the torque applied by the environment to the follower.
When the control objects \eqref{math5:theta} and \eqref{math5:tau} are satisfied, the leader and follower behave as if they are connected by a rigid rod, as shown in the right-hand figure of Fig.~\ref{fig:bilateralctrl}.

In this study, PD control was used for position control, and P control was used for force control.
The discrete-time position controller of the leader is
\begin{align}
  e_{k+1} &= \theta_{fk} - \theta_{lk} \\
  \dot{e}_{k+1} &= \frac{-2g_p}{2+g_p T_s} e_k + \frac{2-g_p T_s}{2+g_p T_s} \dot{e}_k + \frac{2g_p}{2+g_p T_s} (\theta_{fk} - \theta_{lk}) \\
  \ddot{\theta}^{ref}_{k} &= -\frac{2 K_{d} g_p}{2 + g_p T_s} e_k + \frac{K_{d} (2 - g_p T_s)}{2 + g_p T_s} \dot{e}_k \\
  &\quad+ \left(K_{p} + \frac{2 K_{d} g_p}{2 + g_p T_s}\right) (\theta_{fk} - \theta_{lk})
\end{align}
where $e$ and $\dot{e}$ are the state variables, $K_{p}$ and $K_{d}$ are the P and D gains, respectively, and $g_p$ is the cut-off frequency of the low-pass filter of the position controller.

The discrete-time DOB and RFOB are given by bi-linear transformations \eqref{math3:dob} and \eqref{math4:tauext2}.
\begin{align}
  q_{k+1} &= A_d q_k + B_{d1} i_k + B_{d2} \omega_k \label{math5:ddob} \\
  \hat{\tau}^{d}_k &= C_{d} q_k + D_{d1} i_k + D_{d2} \omega_k \label{math5:tauhatd} \\
  z_{k+1} &= A_r z_k + B_{r1} i_k + B_{r2} \omega_k + B_{r3} f(\theta, \omega) \label{math5:drfob} \\
  \hat{\tau}^{e}_k &= C_{r} z_k + D_{r1} i_k + D_{r2} \omega_k + D_{r3} f(\theta, \omega)
  \label{math5:tauexthatd}
\end{align}
In addition, the leader force controller is
\begin{align}
  \tau^{ref}_k &= K_f (\hat{\tau}^{e}_{f(k-1)} + \hat{\tau}^{e}_{l(k-1)})
\end{align}
where $K_f$ denotes the P gain of the force controller.
Hence, control input $i$ of the leader is
\begin{align}
  i_{k+1} &= \frac{\bar{J}}{2 K}\ddot{\theta}^{ref}_k + \frac{1}{K} \hat{\tau}^{d}_k - \frac{K_f}{2 K} \tau^{ref}_k \\
  &= \frac{\bar{J}}{2K} \frac{-2 K_{d} g_p}{2 + g_p T_s} e_k + \frac{\bar{J}}{2K} \frac{K_{d} (2 - g_p T_s)}{2 + g_p T_s} \dot{e}_k \\
  &\quad + \frac{\bar{J}}{2K} \left(K_{p} + \frac{2 K_{d} g_p}{2 + g_p T_s}\right) (\theta_{fk} - \theta_{lk}) \\
  &\quad + \frac{C_{d}}{K} q_k + \frac{D_{d1}}{K} i_k + \frac{D_{d2}}{K} \omega_k \\
  &\quad - \frac{K_f}{2 K} (\hat{\tau}^{e}_{f(k-1)} + \hat{\tau}^{e}_{l(k-1)}).\label{math:i}
\end{align}

Based on the above, the state-space representation of the leader controller is given as follows:
\begin{equation}
  \left\{ \,
    \begin{aligned}
      x_{l(k+1)} &= A_c x_{lk} + B_c v_{lk} \\
      u_{lk} &= C_c x_{lk} + D_c v_{lk}
    \end{aligned}
  \right.
  \label{math:lcontroller}
\end{equation}
\begin{align}
  A_c &=
  \begin{bmatrix}
    0 & 0 & 0 & 0 & 0 \\
    A_{21} & A_{22} & 0 & 0 & 0 \\
    0 & 0 & A_{d1} & 0 & B_{d1} \\
    0 & 0 & 0 & A_{r1} & B_{r1} \\
    A_{51} & A_{52} & \frac{C_{d}}{K} & 0 & \frac{D_{d1}}{K}
  \end{bmatrix}, \\
  A_{c21} &= \frac{-2 K_{d} g_p}{2 + g_p T_s},~
  A_{c22} = \frac{2-g_p T_s}{2+g_p T_s}, \\
  A_{c51} &= \frac{\bar{J}}{2K} \frac{-2 K_{d} g_p}{2 + g_p T_s},~
  A_{c52} = \frac{\bar{J}}{2K} \frac{K_{d} (2 - g_p T_s)}{2 + g_p T_s}, \\
	B_c &=
  \begin{bmatrix}
    1 & -1 & 0 & 0 & 0 & 0 \\
    B_{c21} & B_{c22} & 0 & 0 & 0 & 0 \\
    0 & 0 & B_{d2} & 0 & 0 & 0 \\
    0 & 0 & B_{r2} & 0 & 0 & B_{r3} \\
    B_{c51} & B_{c52} & \frac{D_{d2}}{K} & - \frac{K_f}{2 K} & - \frac{K_f}{2 K} & 0
  \end{bmatrix}, \\
  B_{c21} &= -B_{c22} = \frac{2g_p}{2+g_p T_s}, \\
  B_{c51} &= -B_{c52} = \frac{\bar{J}}{2K} \left(K_{p} + \frac{2 K_{d} g_p}{2 + g_p T_s}\right) ,\\
  C_c &=
  \begin{bmatrix}
    0 & 0 & C_{d} & 0 & D_{d1} \\
    0 & 0 & 0 & C_{r} & D_{r1}
  \end{bmatrix}, \\
  D_c &=
  \begin{bmatrix}
    0 & 0 & D_{d2} & 0 & 0 & 0 \\
    0 & 0 & D_{r3} & 0 & 0 & D_{r3}
  \end{bmatrix}, \\
  x_l &=
  \begin{bmatrix}
    e \\
    \dot{e} \\
    q \\
    z \\
    i
  \end{bmatrix},~
  u_l =
  \begin{bmatrix}
    \hat{\tau}^{d} \\
    \hat{\tau}^{e}_l
  \end{bmatrix},~
  v_l =
  \begin{bmatrix}
    \theta_f \\
    \theta_l \\
    \omega \\
    \hat{\tau}^{e}_f \\
    \hat{\tau}^{e}_l \\
    f(\theta, \omega)
  \end{bmatrix}.
\end{align}
The state-space representation of the follower controller is the replacement of $\theta_l,~\theta_f$ and $\hat{\tau}^{e}_l,~\hat{\tau}^{e}_f$ in the state-space representation of the leader controller. 
\begin{equation}
  \left\{ \,
    \begin{aligned}
      x_{f(k+1)} &= A_c x_{fk} + B_c v_{fk} \\
      u_{fk} &= C_c x_{fk} + D_c v_{fk}
    \end{aligned}
  \right.
  \label{math:fcontroller}
\end{equation}
\begin{align}
  x_f &=
  \begin{bmatrix}
    e \\
    \dot{e} \\
    q \\
    z \\
    i
  \end{bmatrix},~
  u_f =
  \begin{bmatrix}
    \hat{\tau}^{d} \\
    \hat{\tau}^{e}_f
  \end{bmatrix},~
  v_f =
  \begin{bmatrix}
    \theta_l \\
    \theta_f \\
    \omega \\
    \hat{\tau}^{e}_l \\
    \hat{\tau}^{e}_f \\
    f(\theta, \omega)
  \end{bmatrix}, \\
\end{align}
 which correspond to the red and blue dotted lines in Fig.~\ref{fig:bd_4ch_a}, respectively.

%%%
\section{Encrypted four-channel bilateral control system\label{sec:encryption}}
This section discusses the encryption of the leader and follower controllers of the four-channel bilateral control system shown in Fig.~\ref{fig:bd_4ch_a} by using the ElGamal encryption\cite{elgamal1985}.
Section~\ref{sec:encctrl} introduces encrypted control using the ElGamal encryption, and Section~\ref{sec:impl} describes the secure implementation of controllers to obtain an encrypted four-channel bilateral control system.

\subsection{Preliminary for encrypted control\label{sec:encctrl}}
The ElGamal encryption $\mathcal{E}=(\mathsf{Gen},\mathsf{Enc},\mathsf{Dec})$ consists of three algorithms: key generation $\mathsf{Gen}$, encryption $\mathsf{Enc}$, and decryption $\mathsf{Dec}$~\cite{elgamal1985}.
$\mathsf{Gen}$ generates a public key $k_p$ and private key $k_s$ on input with a key length of ${\lambda}\ \si{bit}$: $\mathsf{Gen}(1^\lambda) = (k_p, k_s).$
$\mathsf{Enc}$ encrypts the plaintext $m$ using public key $k_p$: $\mathsf{Enc}(k_p, m) = c$, where $c$ is the ciphertext.
$\mathsf{Dec}$ decrypts the ciphertext $c$ using the public key $k_p$ and the secret key $k_s$: $\mathsf{Dec}(k_s, c) = m'$.
Note that $\mathsf{Dec}(k_s, \mathsf{Enc}(k_p, m)) = m$ holds.
ElGamal encryption is a multiplicative homomorphic encryption scheme that allows ciphertext multiplication.
Let $c_1 = \mathsf{Enc}(k_p, m_1)$, $c_2 = \mathsf{Enc}(k_p, m_2)$.
The ciphertext of product $m_1$ $m_2$ is $\mathsf{Enc}(k_p, m_1 m_2) = c_1 \times_e c_2$, where $\times_e$ denotes the element-wise product.
This property is called multiplicative homomorphism.

Because the ElGamal encryption can encrypt only the member in plaintext space $\mathcal{M}$, a conversion from real numbers to plaintext space $\mathbb{R} \rightarrow \mathcal{M}$ is necessary.
The conversion $\mathbb{R} \rightarrow \mathcal{M}$ from real numbers to plaintext space is as follows:
\begin{align}
    \mathbb{R} \ni x \mapsto \bar{x} = \lceil \gamma x + a \rfloor \in \mathcal{M},~
    a \coloneqq \left\{
    \begin{array}{l}
    0\ \mathrm{if}\ x \geq 0, \\
    p\ \mathrm{if}\ x < 0.
    \end{array}
    \right.
\end{align}
Note that $\gamma$ is the appropriate scaling parameter and $\lceil \cdot \rfloor$ is the rounding to the nearest member in the plaintext space $\mathcal{M}$.
This transformation results in a quantization error.

\subsection{Secure implementation of controllers\label{sec:impl}}
\begin{figure*}[tb]
  \centering
  \subfloat[]{\includegraphics[scale=.98]{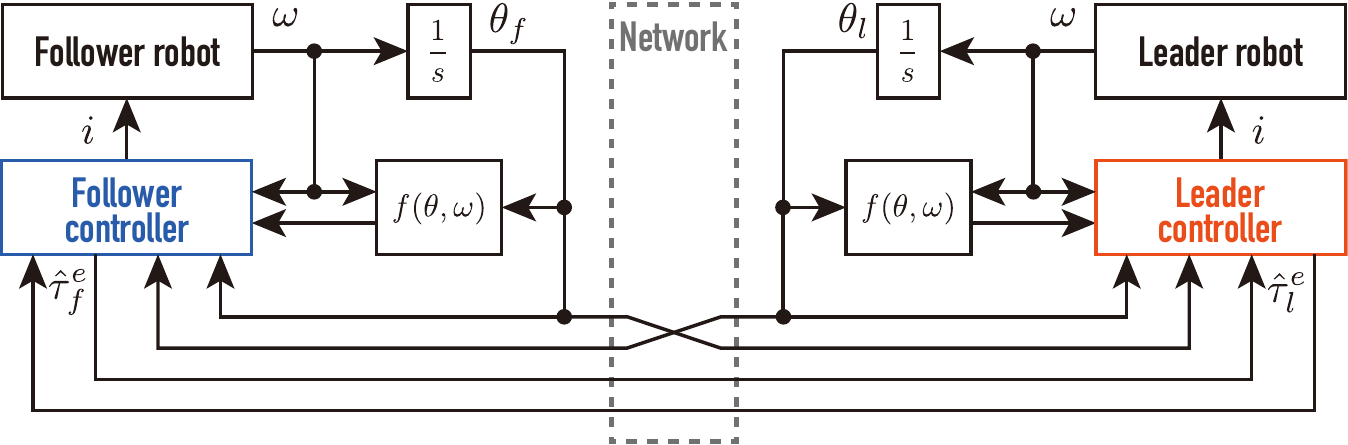}%
  \label{fig:bd_4ch_b}} \\
  \subfloat[]{\includegraphics[scale=.98]{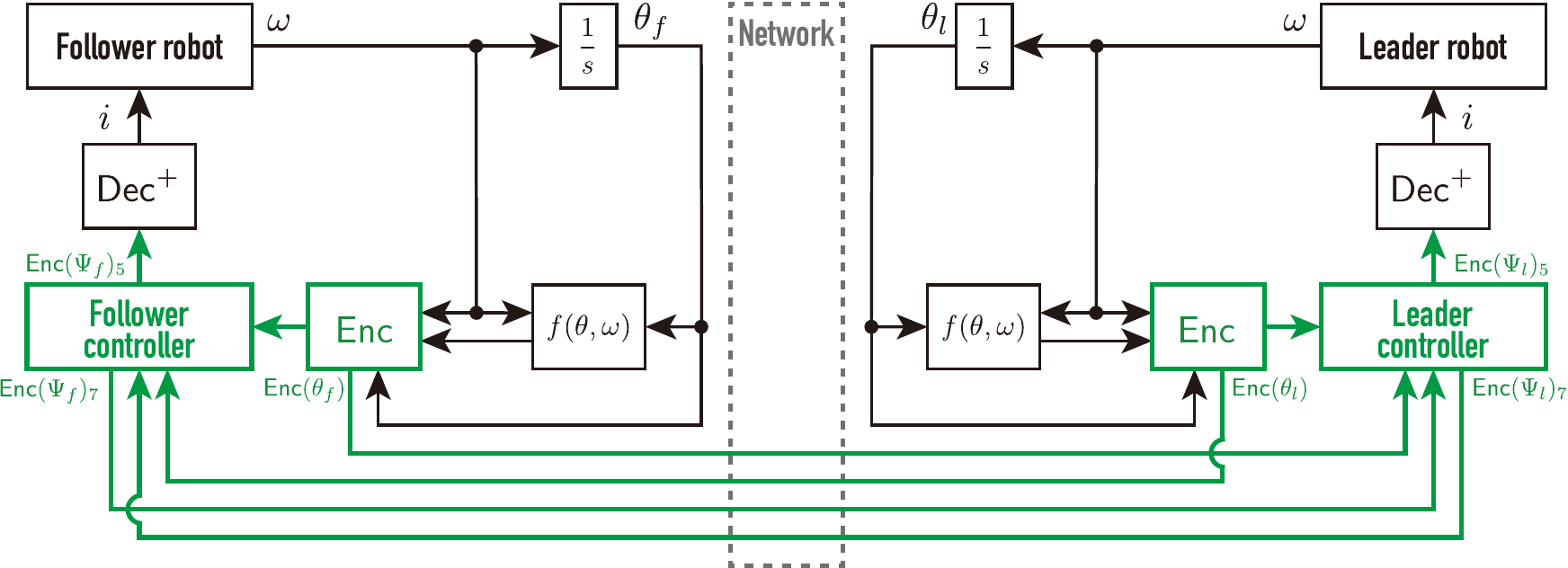}%
  \label{fig:bd_4ch_c}}
  \caption{Block diagrams of unencrypted and encrypted four-channel bilateral control systems. The follower and leader controllers in (a) are securely implemented to obtain the proposed control system in (b), where the green lines and blocks represent encrypted communication and homomorphic operations, respectively.}
  \label{fig:bd_4ch}
\end{figure*}

This section describes an encrypted four-channel bilateral control system that uses a controller encryption technique~\cite{Kogiso2015}.
Fig.~\ref{fig:bd_4ch}(a) shows an entire block diagram of the unencrypted four-channel bilateral control system, where the leader and follower controllers are described in red and blue blocks, respectively.

The procedure for implementing the encrypted leader and follower controllers is as follows.
First, the leader and follower controllers~\eqref{math:lcontroller} and \eqref{math:fcontroller} are transformed as follows:
\begin{align}
  \psi_{lk} &= \Phi_l \xi_{lk} \coloneqq f_l(\Phi_l, \xi_{lk}) \label{math:lctrlr},\ 
  \psi_{lk} \coloneqq
  \begin{bmatrix}
    x_{l(k+1)} \\
    u_{lk}
  \end{bmatrix} \in \mathbb{R}^{7},\\[1ex]
  \Phi_l &\coloneqq
  \begin{bmatrix}
    A_c & B_c \\
    C_c & D_c
  \end{bmatrix} \in \mathbb{R}^{7\times11},\ 
  \xi_{lk} \coloneqq
  \begin{bmatrix}
    x_{lk} \\
    v_{lk}
  \end{bmatrix} \in \mathbb{R}^{11}, \label{math:xil}\\
  \psi_{fk} &= \Phi_f \xi_{fk} \coloneqq f_f(\Phi_f, \xi_{fk}) \label{math:fctrlr},\ 
  \psi_{fk} \coloneqq
  \begin{bmatrix}
    x_{f(k+1)} \\
    u_{fk}
  \end{bmatrix} \in \mathbb{R}^{7},\\[1ex]
  \Phi_f &= \Phi_l,\ 
  \xi_{fk} \coloneqq
  \begin{bmatrix}
    x_{fk} \\
    v_{fk}
  \end{bmatrix} \in \mathbb{R}^{11}
\end{align}
where $f_l$ and $f_f$ are the leader and follower controllers, respectively, and $\Phi_l$ and $\Phi_f$ are the controller parameters of the leader and follower, respectively.
Let $f_l = f_l^{+} \circ f_l^{\times}$, where $f_l^{+}$ and $f_l^{\times}$ denote addition and multiplication on the leader, respectively, and similarly, $f_f=f_f^{+} \circ f_f^{\times}$ is defined on the follower. 
If the quantization error is negligibly small, the leader and follower encrypted controllers $f^{\times}_{l\mathcal{E}^+}$ and $f^{\times}_{f\mathcal{E}^+}$ are represented as follows:
\begin{subequations}
\begin{align}
    f^{\times}_{l\mathcal{E}^+} &: (\mathsf{Enc}(\bar{\Phi}_l),~\mathsf{Enc}(\bar{\xi}_{lk})) \mapsto \mathsf{Enc}({\Psi}_{lk}), 
    \label{math:enclc} \\
    f^{\times}_{f\mathcal{E}^+} &: (\mathsf{Enc}(\bar{\Phi}_f),~\mathsf{Enc}(\bar{\xi}_{fk})) \mapsto \mathsf{Enc}({\Psi}_{fk}),
    \label{math:encfc}
\end{align}
\end{subequations}
where $\Psi_{lk} \coloneqq f_l^\times(\Phi_l, \xi_{lk})$ and $\Psi_{fk} \coloneqq f_l^\times(\Phi_f, \xi_{fk})$, which are computed using the multiplicative homomorphism.
The bar indicates quantization.
Define $\mathsf{Dec}^+ \coloneqq f^+ \circ \mathsf{Dec}$. 
Then, $\bar{\psi}_{lk}$ and $\bar{\psi}_{fk}$ are obtained as follows:
\begin{align}
    \mathsf{Dec}^+(\mathsf{Enc}(\Psi_{lk})) &= {\psi}_{lk},\quad
    \mathsf{Dec}^+(\mathsf{Enc}(\Psi_{fk})) = {\psi}_{fk}.
\end{align}
Let $\mathcal{E}^+ \coloneqq (\mathsf{Gen}, \mathsf{Enc}, \mathsf{Dec}^+)$ be a modified ElGamal encryption scheme.

Next, to realize $f^{\times}_{l\mathcal{E}^+}$, we provide the details of the encrypted leader controller \eqref{math:enclc} and highlight it on the leader side in Fig.~\ref{fig:bd_4ch}(b).
The arguments of $f^{\times}_{l\mathcal{E}^+}$ are realized as follows:
1) $\mathsf{Enc}(\bar{\Phi}_l)$: Quantized and encrypted parameters of the leader controller.
2) $\mathsf{Enc}(\bar{\xi}_{lk})$: Quantized and encrypted leader controller inputs $\xi_{l}$.
From the definition of $\xi_l$ in \eqref{math:xil}, it contains the controller state variable $x_{l}$ and the plant output $v_{l}$.
$v_{l}$ must be encrypted because $x_{l}$ is kept in the ciphertext of the controller.
The definition of $v_l$ in the leader controller \eqref{math:lcontroller} contains signals $\theta_f$, $\theta_l$, $\omega$, $\hat{\tau}^e_f$, $\hat{\tau}^e_l$, and $f(\theta, \omega)$. 
Because the ciphertexts of $\theta_f$ and $\hat{\tau}^e_f$ are sent from the follower, the other signals, i.e., $\theta_l$, $\omega$, $\hat{\tau}^e_l$, and $f(\theta,\,\omega)$, must be encrypted before inputting them to the leader controller.
Moreover, \eqref{math:enclc} returns $\mathsf{Enc}(\bar{\Psi}_{lk})$, which includes the ciphertexts of the estimated reaction force $\hat{\tau}^e_l$ and input current $i$.
$\mathsf{Enc}(\Psi_l)_7$ and $\mathsf{Enc}(\Psi_l)_5$ correspond to $\hat{\tau}^e_l$ and $i$, respectively, 
where $\mathsf{Enc}(\Psi_l)_j$ denotes the $j$th element of the control input ciphertext $\mathsf{Enc}(\Psi_l)$.
$\mathsf{Enc}(\Psi_l)_7$ is sent to the follower controller with $\mathsf{Enc}(\theta_l)$, and $\mathsf{Enc}(\Psi_l)_5$ is sent to $\mathsf{Dec}^+$ to obtain the input current $i$.
At the leader side of the block diagram in Fig.~\ref{fig:bd_4ch}(b), the inputs and outputs of \eqref{math:enclc} correspond to those of the ``Leader controller'' block.
Similarly, $f^{\times}_{f\mathcal{E}^+}$ of \eqref{math:encfc} can be realized via clarifying its arguments and output, which correspond to those of ``Follower controller'' in Fig.~\ref{fig:bd_4ch}(b), respectively.

The features in Fig.~\ref{fig:bd_4ch}(b) are as follows.
Encrypted communication lines and blocks are shown in green.
The leader and follower controller parameters, $\Phi_l$ and $\Phi_f$, respectively, are both encrypted.
Communication signals over the network, i.e., the leader rotation angle $\mathsf{Enc}(\theta_l)$, estimated reaction force $\mathsf{Enc}(\Psi_l)_7$, follower rotation angle $\mathsf{Enc}(\theta_f)$, and estimated reaction force $\mathsf{Enc}(\Psi_f)_7$, are all encrypted.
Thus, the leader and follower controllers, shown in Fig.~\ref {fig:bd_4ch_b}, were implemented securely to realize an encrypted four-channel bilateral control system.

In the next section, teleoperation experiments are presented to validate the proposed encrypted four-channel bilateral control system.

\begin{figure*}[tb]
\centering
\includegraphics[width=1.0\linewidth]{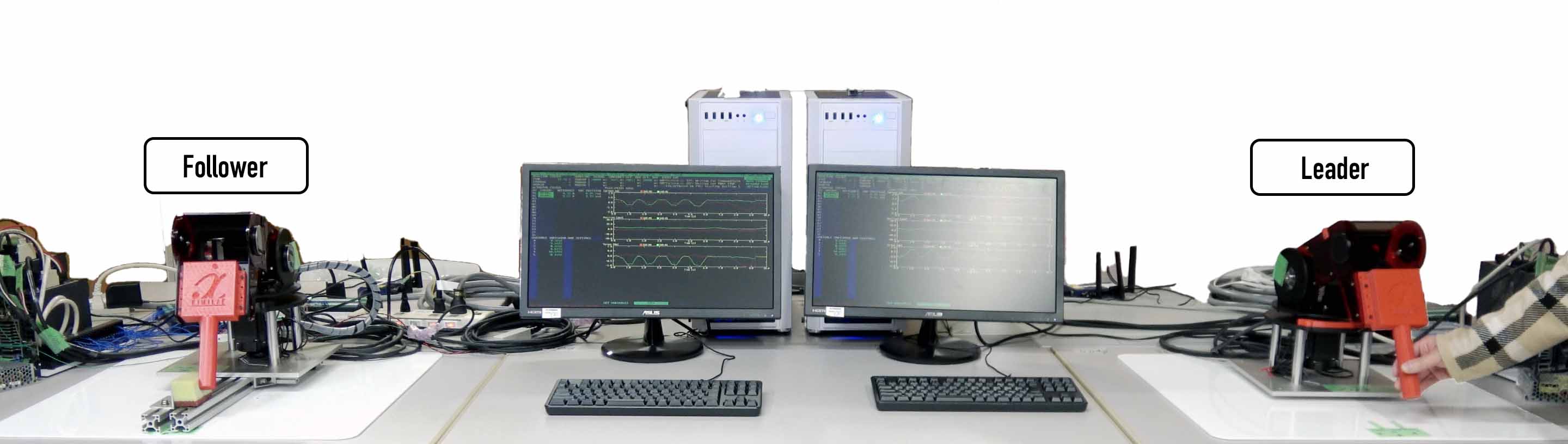}
\caption{Manipulation of leader robot by hand.  \label{fig:setup}}
\end{figure*}

\section{Demonstration of Encrypted four-channel Bilateral Control System\label{sec:experiment}}

This section demonstrates the developed bilateral control system with the leader and follower robot arms and its cyber-secure motion-tracking teleoperation with force feedback.
After the experimental settings are introduced in Section~\ref{sec:es}, the three features of motion-tracking teleoperation, force feedback, and effects of the encryption, are examined in Sections~\ref{sec:free}, \ref{sec:incontact}, and \ref{sec:qerror}, respectively.

%%%
\subsection{Experimental setting}\label{sec:es}

Fig.~\ref{fig:setup} shows the setup used for the demonstration.
First, motion-tracking teleoperation experiments were conducted to confirm that the follower followed the posture of the leader. 
Next, to confirm that the reaction force on the follower side returns to the leader side, an experiment was conducted in which the follower was placed in contact with an object.
In both experiments, the sampling period was $\SI{20}{ms}$ and the cut-off frequency of the pseudo-differentiator was $g_p = \SI{200}{rad/s}$.
TABLE~\ref{tab:four-channelparams} lists the controller parameters and the plant models for each motor.

\begin{figure}[!t]
\centering
\subfloat[]{\includegraphics[scale=.54]{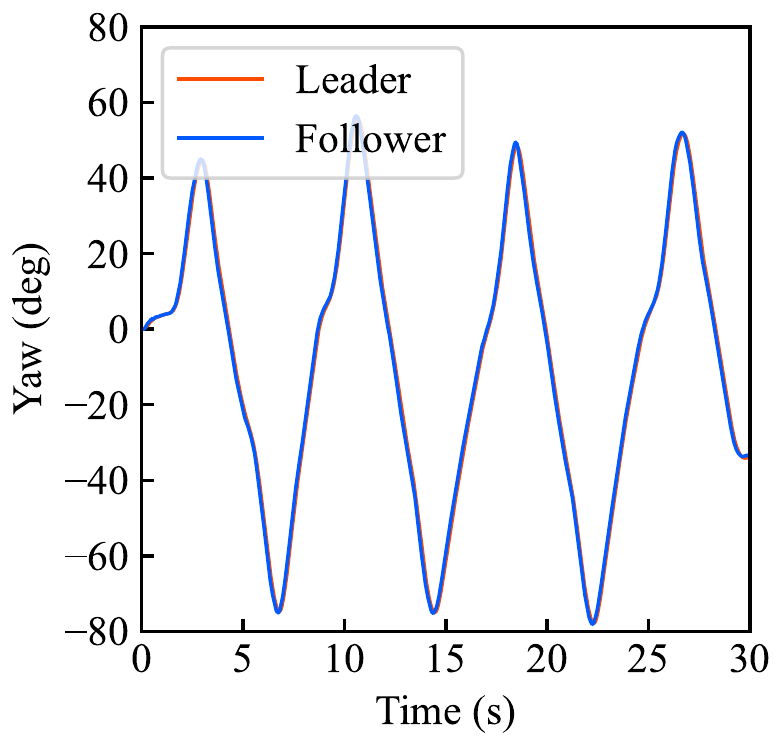}%
\label{fig:angle_four-channely}}
% \hfil
\subfloat[]{\includegraphics[scale=.54]{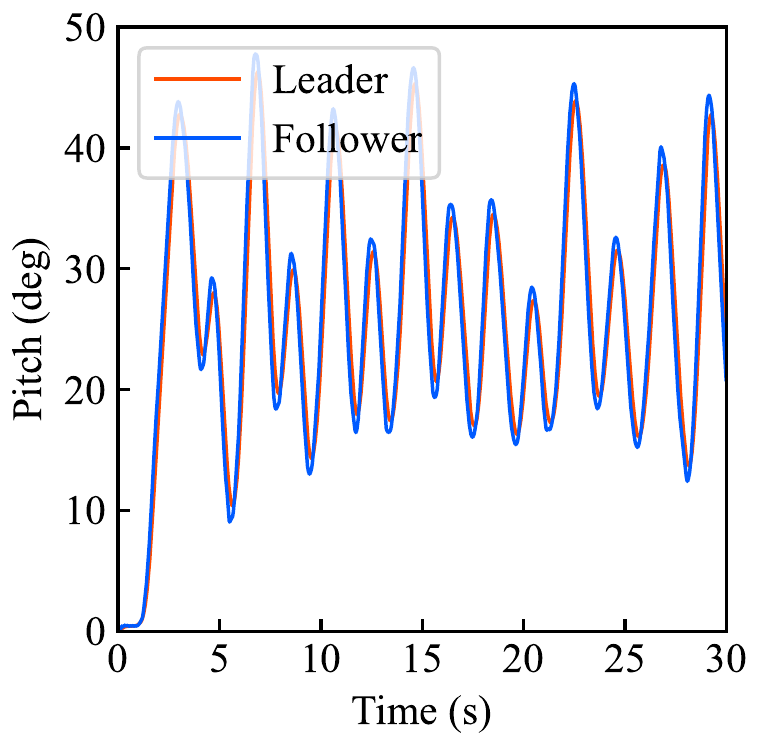}%
\label{fig:angle_four-channelp}} \\
\subfloat[]{\includegraphics[scale=.54]{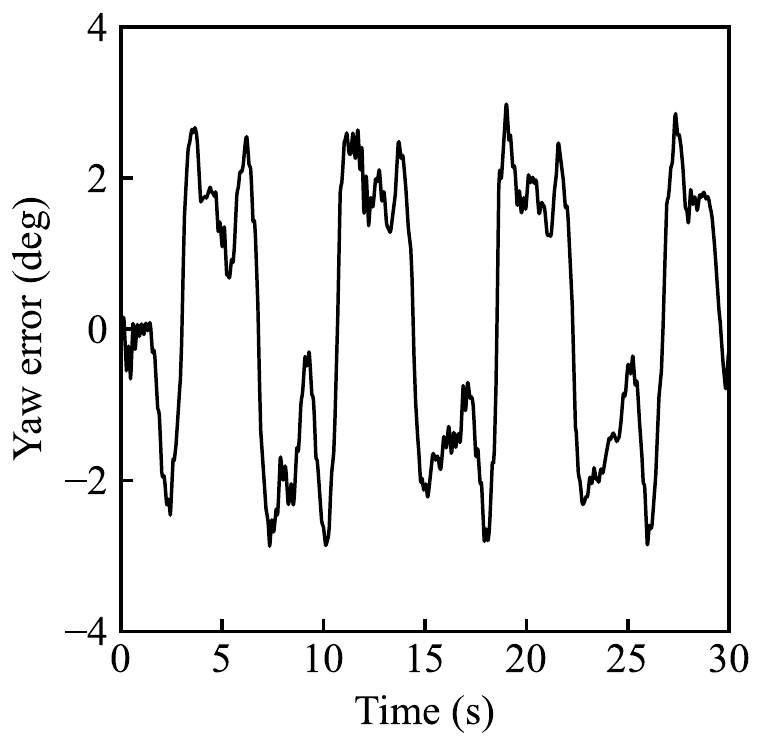}%
\label{fig:errorangle_four-channely}}
% \hfil
\subfloat[]{\includegraphics[scale=.54]{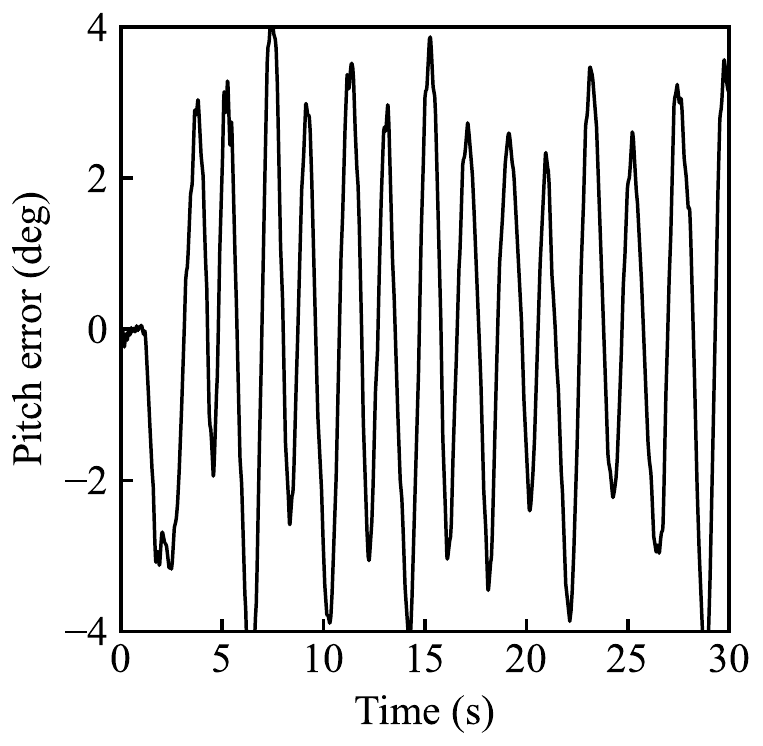}%
\label{fig:errorangle_four-channelp}} \\
\subfloat[]{\includegraphics[scale=.54]{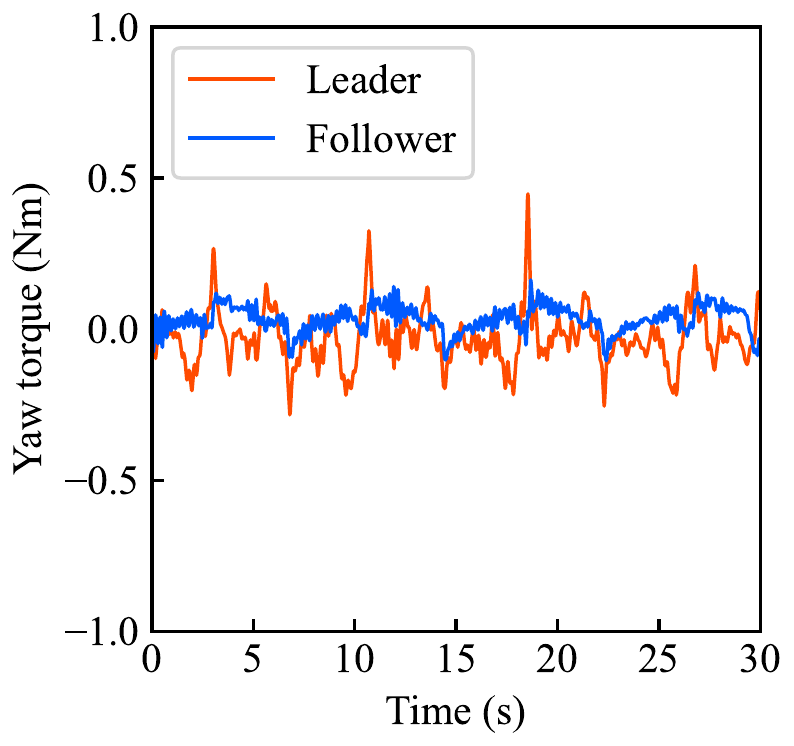}%
\label{fig:tau_four-channely}}
% \hfil
\subfloat[]{\includegraphics[scale=.54]{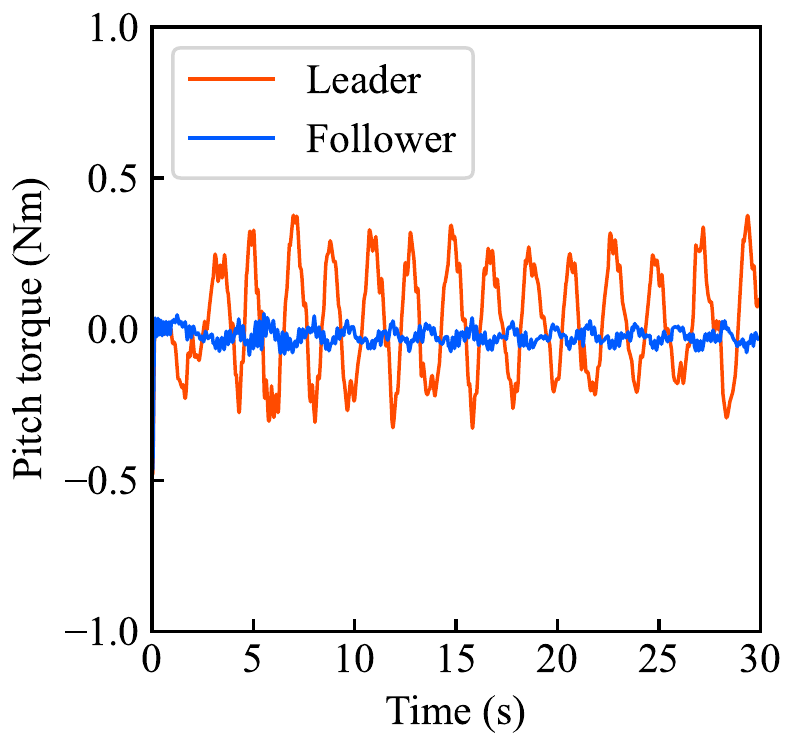}%
\label{fig:tau_four-channelp}}
\caption{Experimental results of unencrypted four-channel bilateral control when no contact with the environment occurs. (a) Rotation angles $\theta_l$ and $\theta_f$ for yaw axis motor. (b) Rotation angles $\theta_l$ and $\theta_f$ for pitch axis motor. (c) Error of angles $\theta_l$ and $\theta_f$ for yaw axis motor. (d) Error of angles $\theta_l$ and $\theta_f$ for pitch axis motor. (e) $\hat{\tau}^{e}_l$ and $\hat{\tau}^{e}_f$: estimated torque regarding the yaw axis motor. (f) $\hat{\tau}^{e}_l$ and $\hat{\tau}^{e}_f$: estimated torque regarding the pitch axis motor.}
\label{fig:four-channelteleope}
\end{figure}

For reference, the results of the teleoperation experiment of the four-channel bilateral control system~\eqref{math:lcontroller} using the parameters listed in TABLE~\ref{tab:four-channelparams} are shown in Figs.~\ref{fig:four-channelteleope}.
Figs.~\ref{fig:four-channelteleope}(a) and (b) show the time responses of the rotation angles of the yaw and pitch axis motors of each of the arm, respectively. 
Figs.~\ref{fig:four-channelteleope}(c) and (d) show the difference between the leader and follower rotation angles during teleoperation.
The figures show the achievement of errors within $\pm \SI{3.0}{deg}$ for the yaw axis and $\pm \SI{4.0}{deg}$ for the pitch axis, confirming that the follower arm follows the behaviors of the leader arm manipulated by the operator. 
The time responses of the estimated reaction torque for each motor are shown in Figs.~\ref{fig:four-channelteleope}(e) and (f), where the red and blue lines represent the leader and follower arms, respectively, and $\hat{\tau}^e_f$ is within $\pm \SI{0.2}{Nm}$.
Therefore, the experimental results confirm that the conventional four-channel bilateral control system achieves motion-tracking teleoperation.

\begin{table}[tb]
\caption{Parameters of the encrypted four-channel bilateral control.}
\label{tab:four-channelparams}
\centering
\begin{tabular}{llll}
\hline
\multicolumn{2}{l}{\textbf{Yaw axis motor}} & \multicolumn{2}{l}{\textbf{Pitch axis motor}} \\
\hline
\multicolumn{4}{l}{\textbf{Plant model}} \\
$\bar{J}$ & ${1.37\times10^{-4}}$\si{kgm^2} & $\bar{J}$ & ${4.1\times10^{-5}}$\si{kgm^2} \\
$K$ & $\SI{0.5}{Nm/A}$   & $K$ & $\SI{0.5106}{Nm/A}$ \\
\hline
\multicolumn{4}{l}{\textbf{Controller parameters}} \\
$K_{p}$ & 300 & $K_{p}$ & 120 \\
$K_{d}$ & 300 & $K_{d}$& 80 \\
$K_{f}$ & 1 & $K_{f}$ & 1 \\
$g_p$ & $\SI{200}{rad/s}$ & $g_p$ & $\SI{200}{rad/s}$ \\
$g_d$ & $\SI{100}{rad/s}$ & $g_d$ & $\SI{100}{rad/s}$ \\
$g_r$ & $\SI{100}{rad/s}$ & $g_r$ & $\SI{100}{rad/s}$ \\
$T_s$ & $\SI{20}{ms}$ & $T_s$ & $\SI{20}{ms}$ \\
\hline
\multicolumn{4}{l}{\textbf{Nonlinear terms model}} \\
$a_{c}$ & 0.0009979 & $a_{g}$ & 0.4696 \\
$b_{c}$ & 0.03152   & $b_{g}$ & 0.01893 \\
$a_{fy}$ & 0.05736  & $c_{g}$ & 1.575 \\
$b_{fy}$ & 12.98    & $a_{fp}$ & 0.009764 \\
$c_{fy}$ & 1.275    & $b_{fp}$ & -1197 \\
$d_{fy}$ & 0.7802   & $c_{fp}$ & 0 \\
	     &          & $d_{fp}$ & 0.05194 \\
	     &          & $e_{fp}$ & 0.003254 \\
\hline
\multicolumn{4}{l}{\textbf{Encryption parameters}} \\
$\lambda$ & $\SI{128}{bit}$ & $\lambda$ & $\SI{128}{bit}$ \\ 
$\gamma_{c}$ & $10^{16}$ & $\gamma_{c}$ & $10^{16}$ \\
$\gamma_{p}$ & $10^{16}$ & $\gamma_{p}$ & $10^{16}$ \\
\hline
\end{tabular}
\end{table}

%%%
\subsection{Motion tracking teleoperation}\label{sec:free}
\begin{figure}[!t]
\centering
\subfloat[]{\includegraphics[scale=.54]{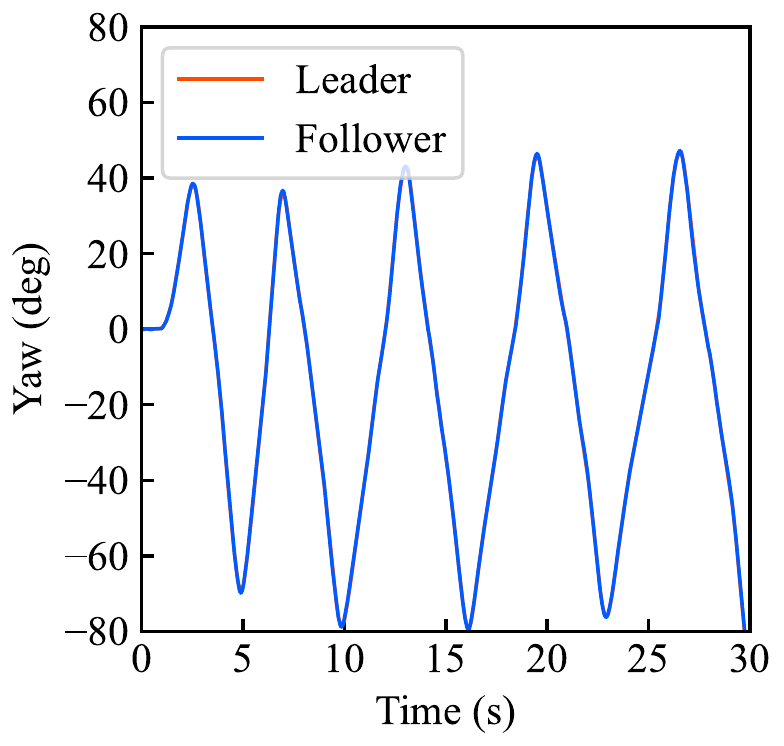}%
\label{fig:angle_encfour-channely}}
% \hfil
\subfloat[]{\includegraphics[scale=.54]{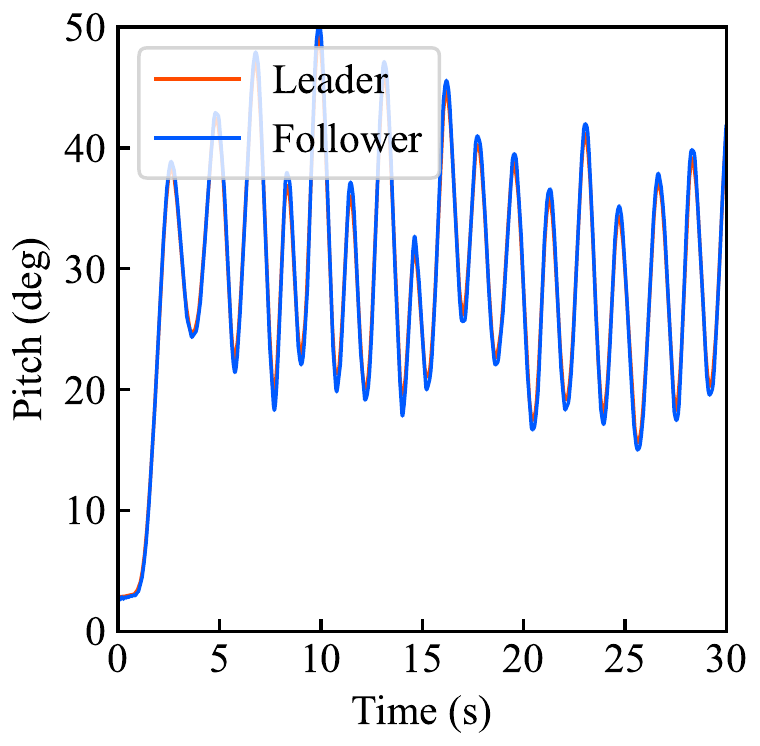}%
\label{fig:angle_encfour-channelp}} \\
\subfloat[]{\includegraphics[scale=.54]{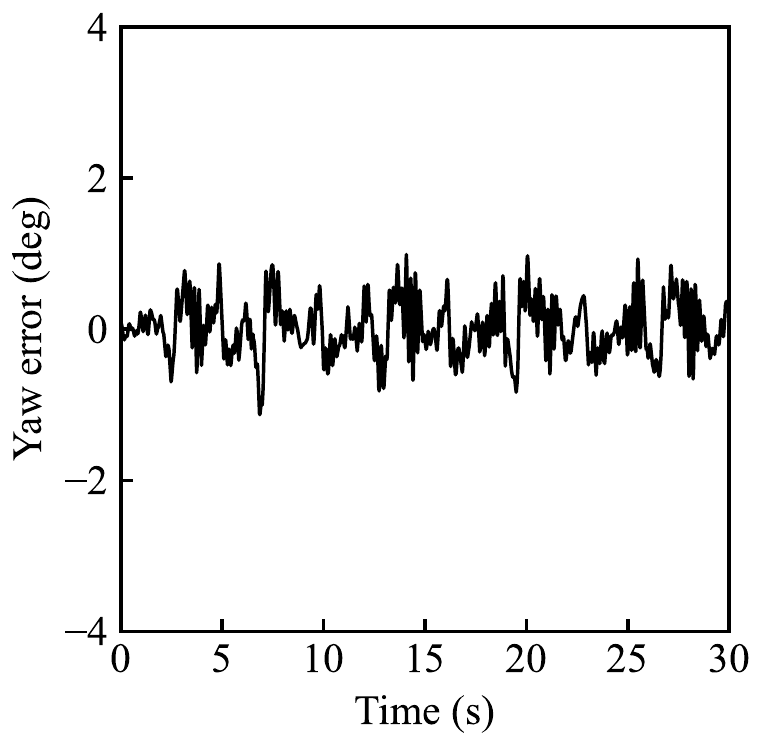}%
\label{fig:errorangle_encfour-channely}}
% \hfil
\subfloat[]{\includegraphics[scale=.54]{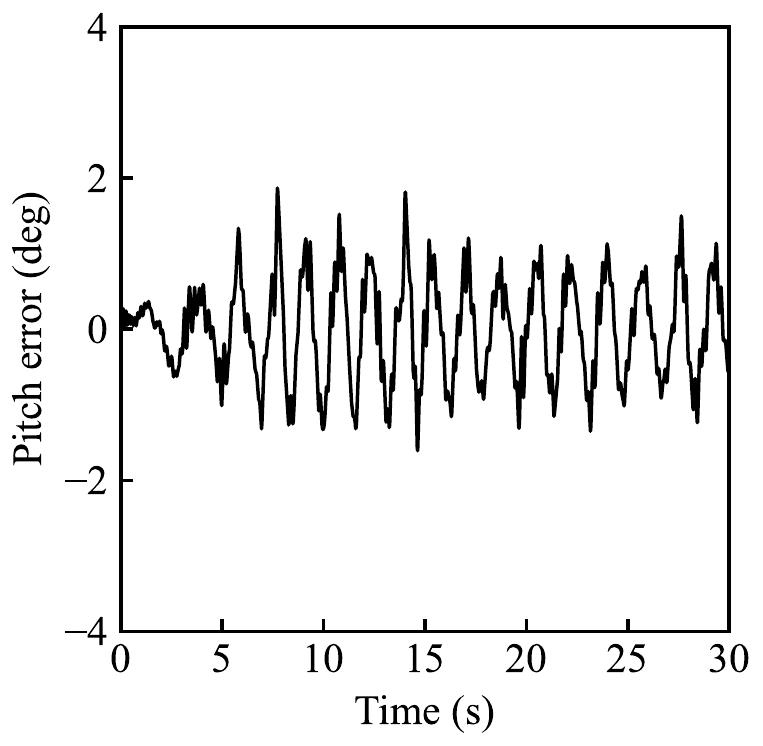}%
\label{fig:errorangle_encfour-channelp}} \\
\subfloat[]{\includegraphics[scale=.54]{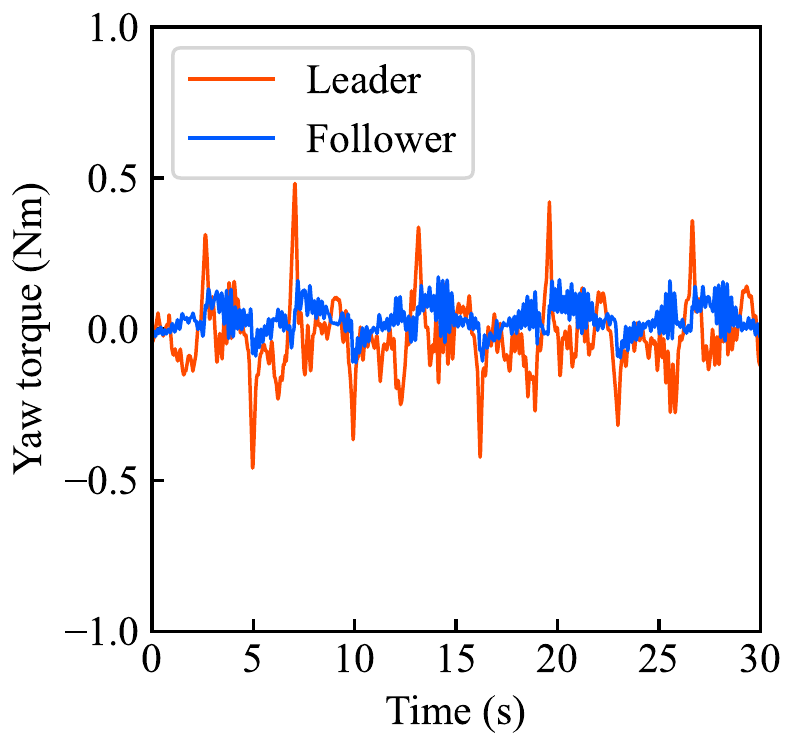}%
\label{fig:tau_encfour-channely}}
% \hfil
\subfloat[]{\includegraphics[scale=.54]{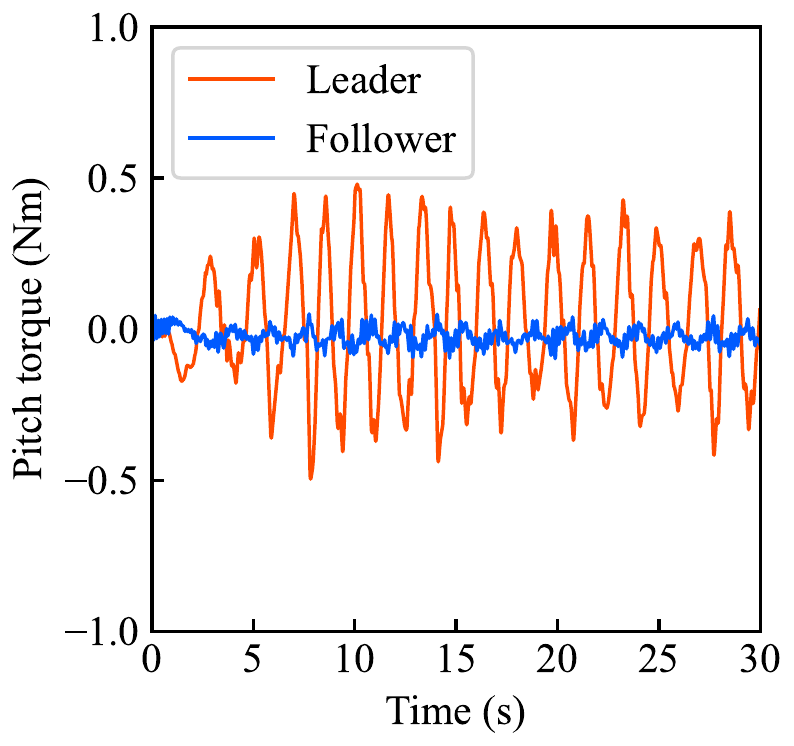}%
\label{fig:tau_encfour-channelp}}
\caption{Experimental results of the developed encrypted bilateral control when there is no contact with the environment. (a) Rotation angles $\theta_l$ and $\theta_f$ for yaw axis motor. (b) Rotation angles $\theta_l$ and $\theta_f$ for pitch axis motor. (c) Error of angles $\theta_l$ and $\theta_f$ for yaw axis motor. (d) Error of angles $\theta_l$ and $\theta_f$ for pitch axis motor. (e) $\hat{\tau}^{e}_l$ and $\hat{\tau}^{e}_f$: Estimated torque for yaw axis motor. (f) $\hat{\tau}^{e}_l$ and $\hat{\tau}^{e}_f$: Estimated torque for pitch axis motor.}
\label{fig:basicteleope}
\end{figure}

This section shows that the angles of each motor of the leader and follower match during teleoperation when an operator operates the leader arm, such that no contact occurs between the follower arm and the environment.
The resulting motion of the robotic arm is shown in Fig.~\ref{fig:basicteleope}.

Figs.~\ref{fig:basicteleope}(a) and (b) show the time responses of the rotation angles regarding the yaw and pitch axis motors of each arm, respectively. 
Figs.~\ref{fig:basicteleope}(c) and (d) show the difference between the leader and follower rotation angles during teleoperation.
The figures show the achievement of errors within $\pm \SI{1.0}{deg}$ for the yaw axis and $\pm \SI{2.0}{deg}$ for the pitch axis, confirming that the follower arm follows the behaviors of the leader arm manipulated by the operator. 
The time responses of the estimated reaction torque for each motor are shown in Figs.~\ref{fig:basicteleope}(e) and (f), where the red and blue lines represent the leader and follower arms, respectively, and $\hat{\tau}^e_f$ is within $\pm \SI{0.2}{Nm}$.
This range is relatively small compared to the case of contact with the environment, which is discussed in \ref{sec:incontact}.
Therefore, the experimental results confirm that the developed control system achieves motion-tracking teleoperation.

%%%
\subsection{Effects of the force feedback}\label{sec:incontact}
\begin{figure}[!t]
\centering
\subfloat[]{\includegraphics[scale=.12]{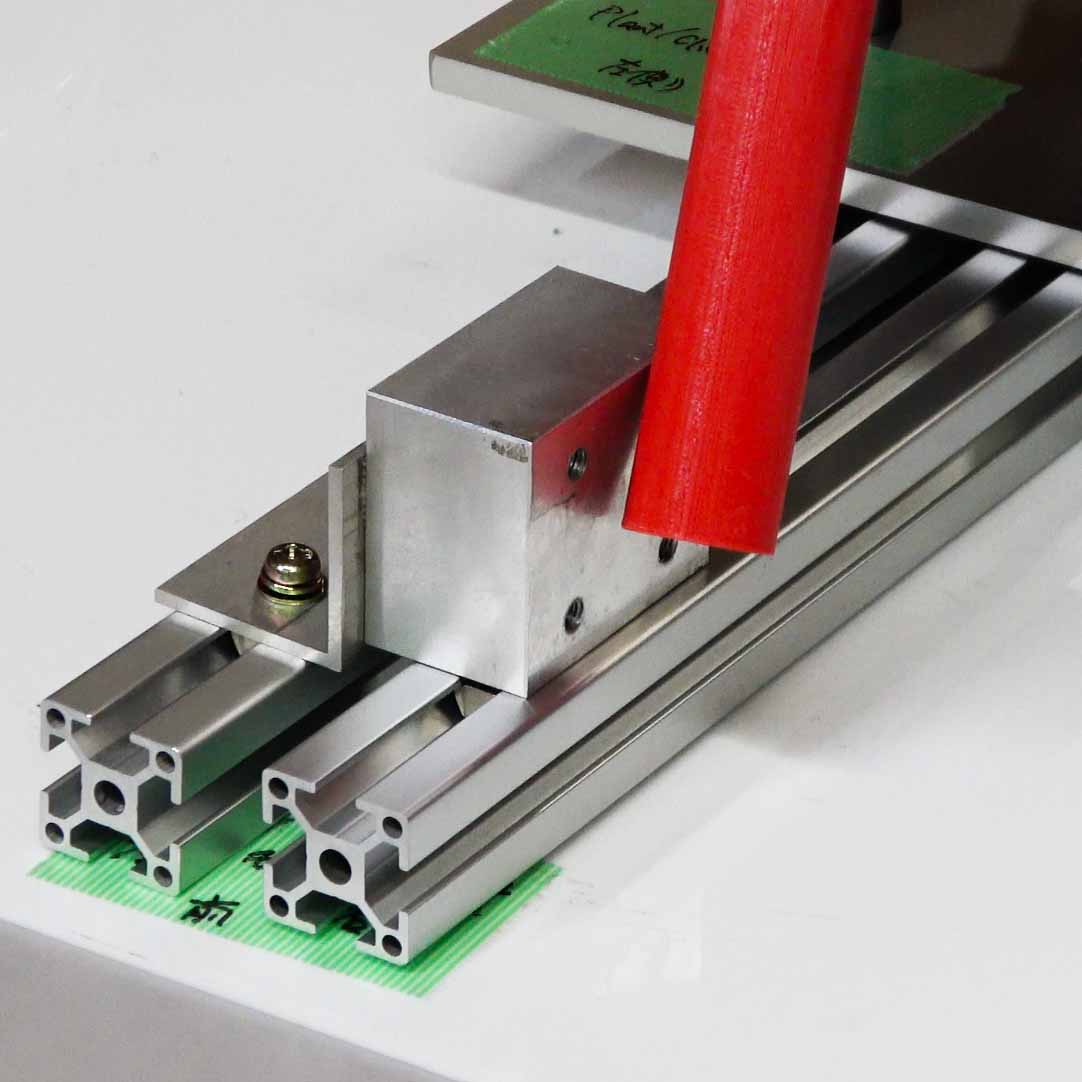}%
\label{fig:aluminum}} \\
\subfloat[]{\includegraphics[scale=.54]{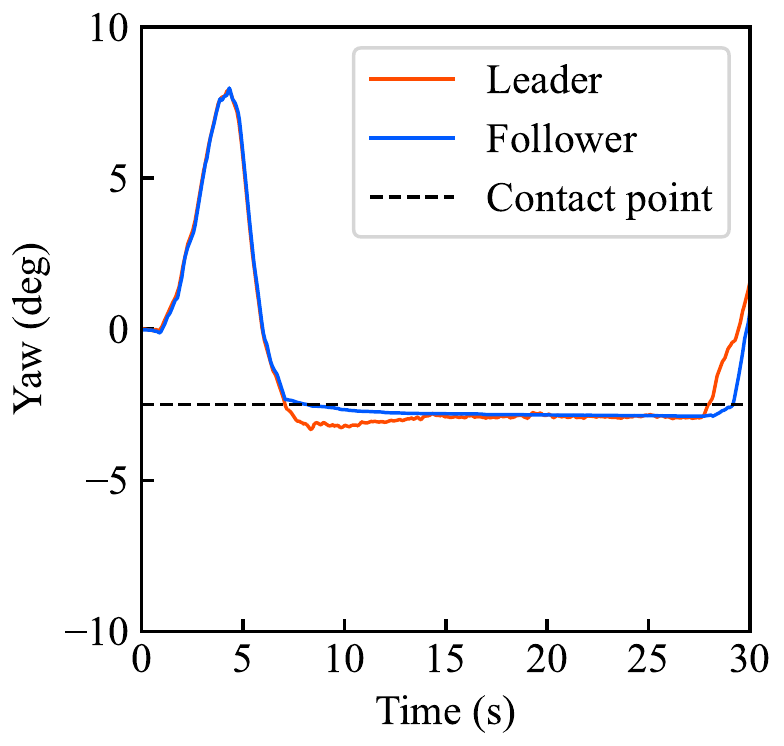}%
\label{fig:angle_encfour-channelay}}
% \hfil
\subfloat[]{\includegraphics[scale=.54]{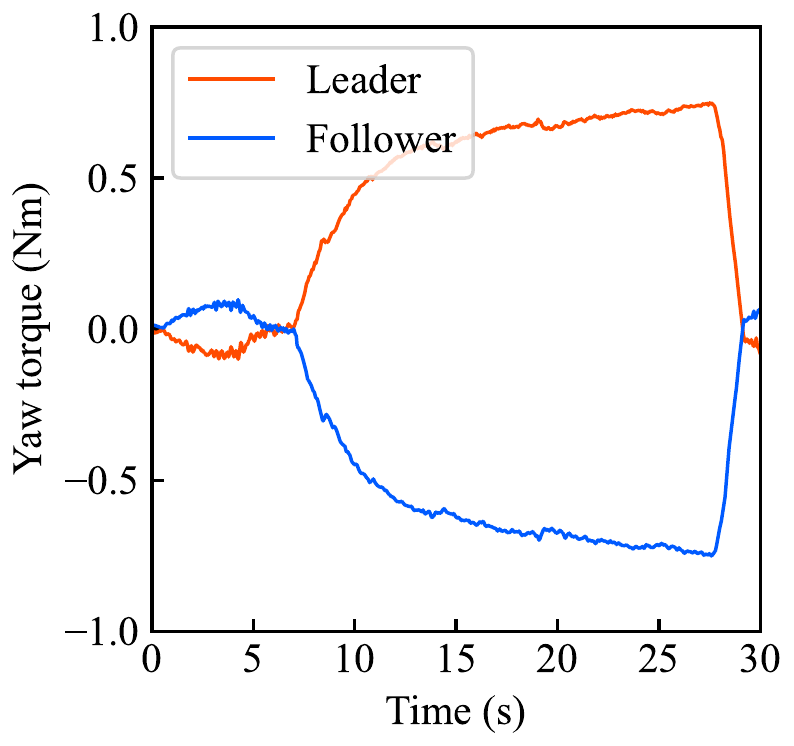}%
\label{fig:tau_encfour-channelay}}
\caption{Experimental results when the robot arm contacts a hard object (aluminum). (a) View of pressing a hand against a hard object (aluminum). (b) $\theta_l$ and $\theta_f$: Rotation angles of yaw axis motor. (c) $\hat{\tau}^{e}_l$ and $\hat{\tau}^{e}_f$: Estimated torque for yaw axis motor.}
\label{fig:incontacthard}
\end{figure}

\begin{figure}[!t]
\centering
\subfloat[]{\includegraphics[scale=.18]{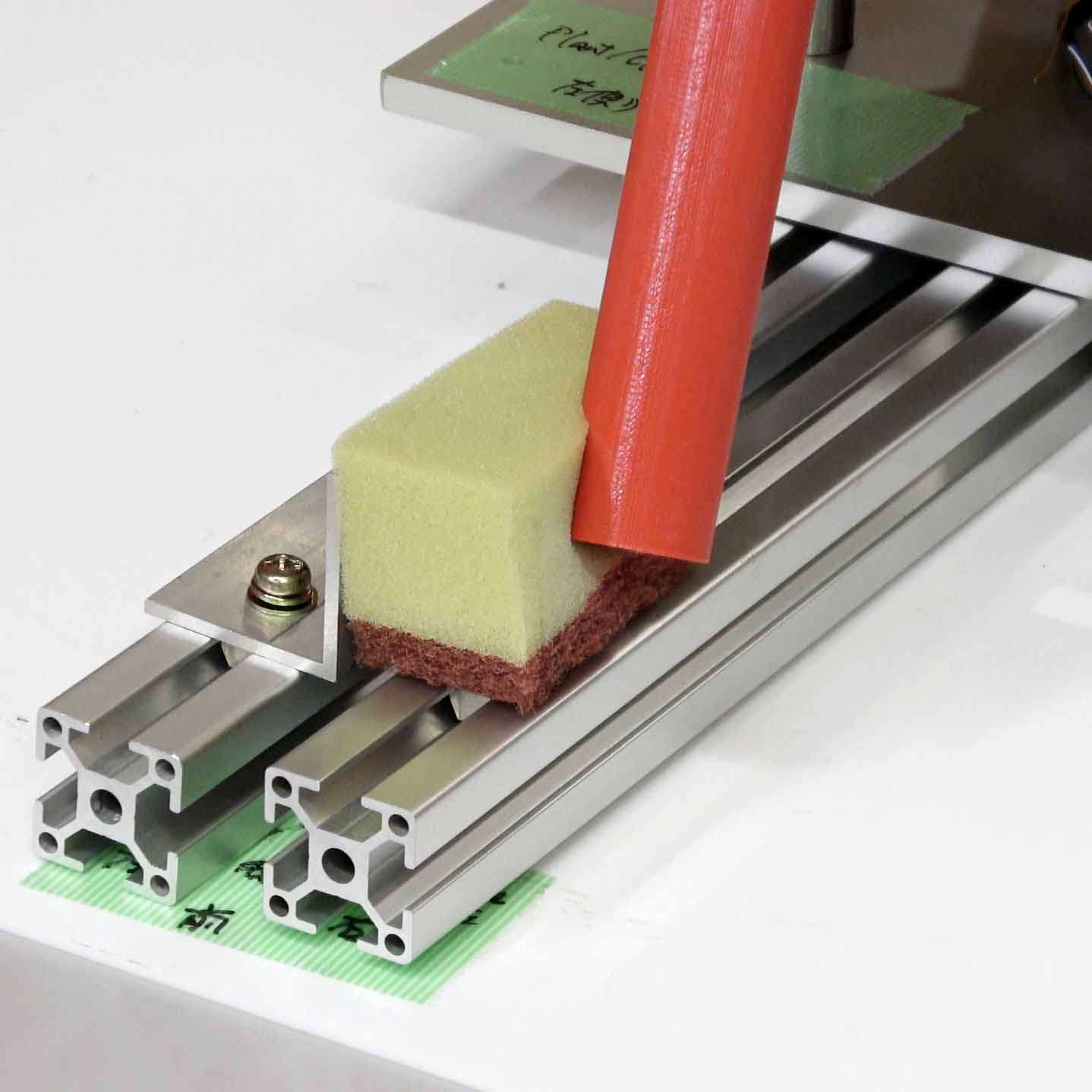}%
\label{fig:sponge}} \\
\subfloat[]{\includegraphics[scale=.54]{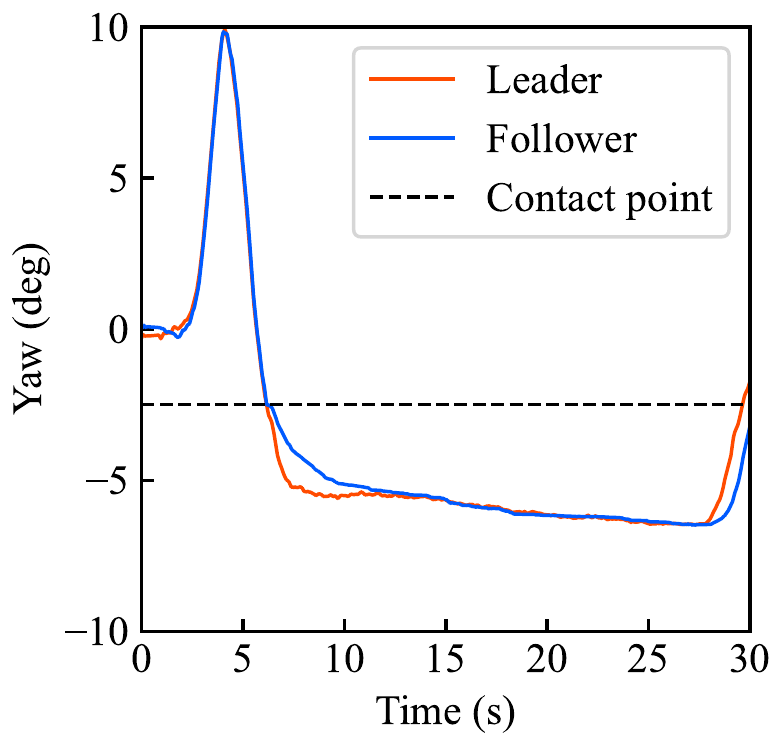}%
\label{fig:angle_encfour-channelsy}}
% \hfil
\subfloat[]{\includegraphics[scale=.54]{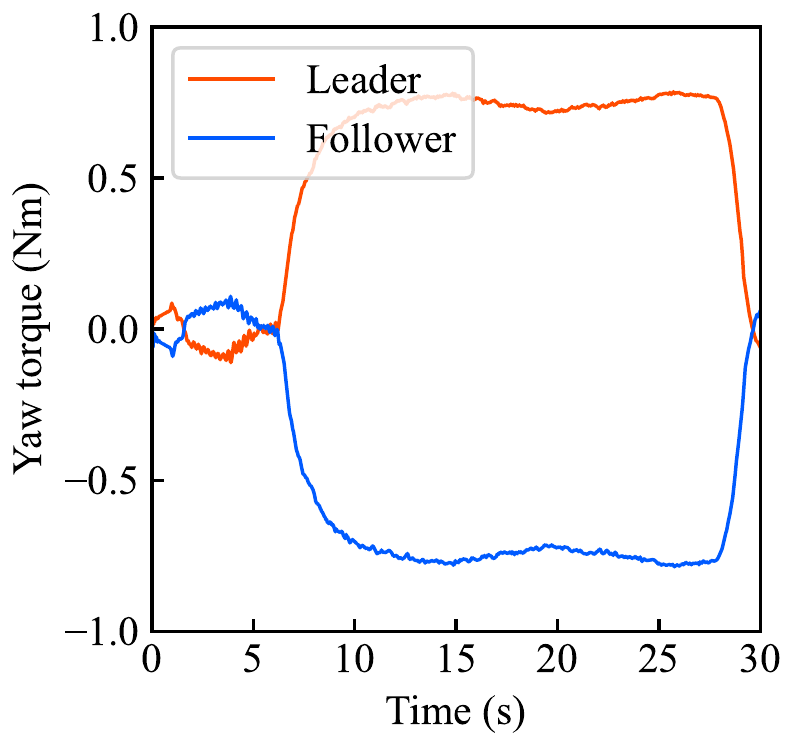}%
\label{fig:tau_encfour-channelsy}}
\caption{Experimental results when the robot arm contacts a soft object (sponge). (a) View of pressing a hand against a soft object (sponge). (b) $\theta_l$ and $\theta_f$: Rotation angles of yaw axis motor. (c) $\hat{\tau}^{e}_l$ and $\hat{\tau}^{e}_f$: Estimated torque for yaw axis motor.}
\label{fig:incontactsoft}
\end{figure}

This section demonstrates that the reaction force applied to the follower returns to the leader when the follower contacts the environment.
The types of environments prepared in this experiment were hard (aluminum block) and soft (sponge).
Figs.~\ref{fig:incontacthard}(a) and \ref{fig:incontactsoft}(a) show the robot arm in contact with aluminum and sponge, respectively.
Figs.~\ref{fig:incontacthard}(b) and \ref{fig:incontactsoft}(b) show the time responses of the rotation angles of the yaw axis motor.
Figs.~\ref{fig:incontacthard}(c) and \ref{fig:incontactsoft}(c) show the time responses of the estimated reaction torque for the yaw axis motor.
In these figures, the red and blue lines represent the leader and follower arms, respectively and the black dotted line represents the starting point of contact with the environment.

Figs.~\ref{fig:incontacthard}(b) and \ref{fig:incontactsoft}(b) confirm that the angle of the follower matches that of the leader.
In addition, Figs.~\ref{fig:incontacthard}(c) and \ref{fig:incontactsoft}(c) confirm that the reaction force applied to the follower returns to that of the leader.
The reaction forces on the leader and follower are opposite in sign and equal in absolute terms. 
This is evident from the control strategy of the four-channel bilateral control~\eqref{math5:tau}, which means that the sum of the reaction forces applied to the leader and follower is zero. 
Moreover, it confirms that in encrypted four-channel bilateral control, differences in the hardness of objects in contact with the follower are transmitted to the leader.
Figs.~\ref{fig:incontacthard}(b) and (c) show that the angles of the leader and follower do not change significantly from the point of contact with aluminum, even when a human applies force to the leader. 
The applied force did not deform the rigid object significantly. 
This indicates that the hardness of aluminum, which the follower is in contact with, is transmitted to the leader.
Figs.~\ref{fig:incontactsoft}(b) and (c) show that when a human applies a force to the leader, the angle between the leader and follower sinks at most $\SI{5}{deg}$ from the point of contact with the sponge. 
Soft objects deform when force is applied. 
This indicates that the softness of the sponge, which the follower is in contact with, is transmitted to the leader.
Therefore, the experimental results confirm that the developed bilateral control system transmits the reaction force on the follower side back to the leader and the difference in the hardness of the environment in which the follower is in contact.

\begin{figure}[!t]
\centering
\subfloat[]{\includegraphics[scale=.54]{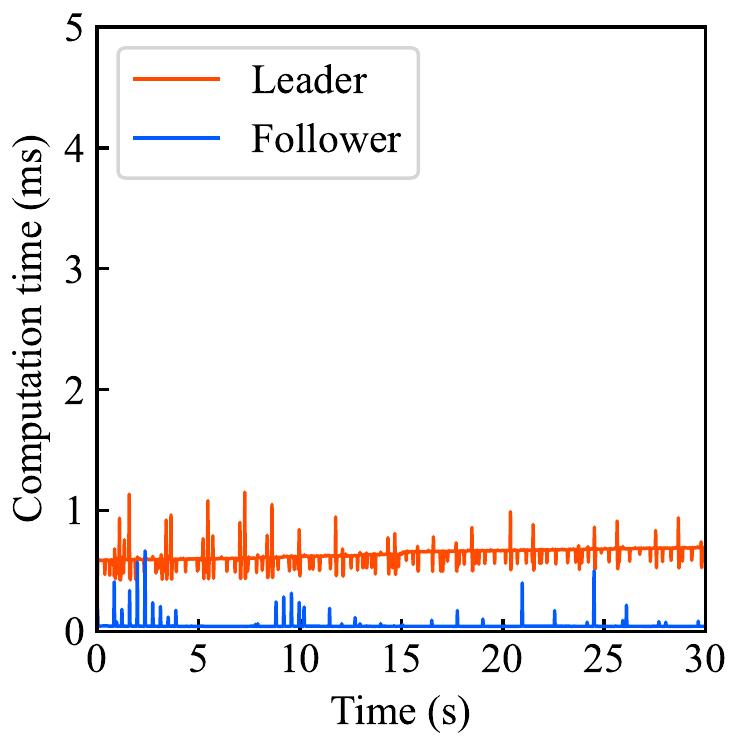}%
\label{fig:tcmp_encfour-channely}}
\subfloat[]{\includegraphics[scale=.54]{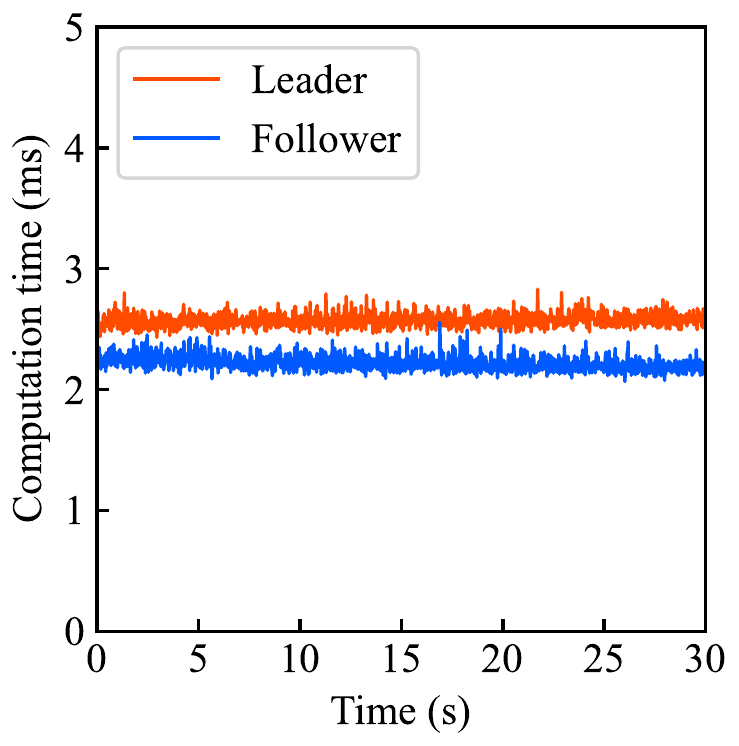}%
\label{fig:tcmp_encfour-channelp}}
\caption{Computation times with and without controller encryption. (a) Computation time without encryption. (b) Computation time with encryption.}
\label{fig:ct}
\end{figure}  

\begin{figure}[!t]
\centering
\subfloat[]{\includegraphics[scale=.54]{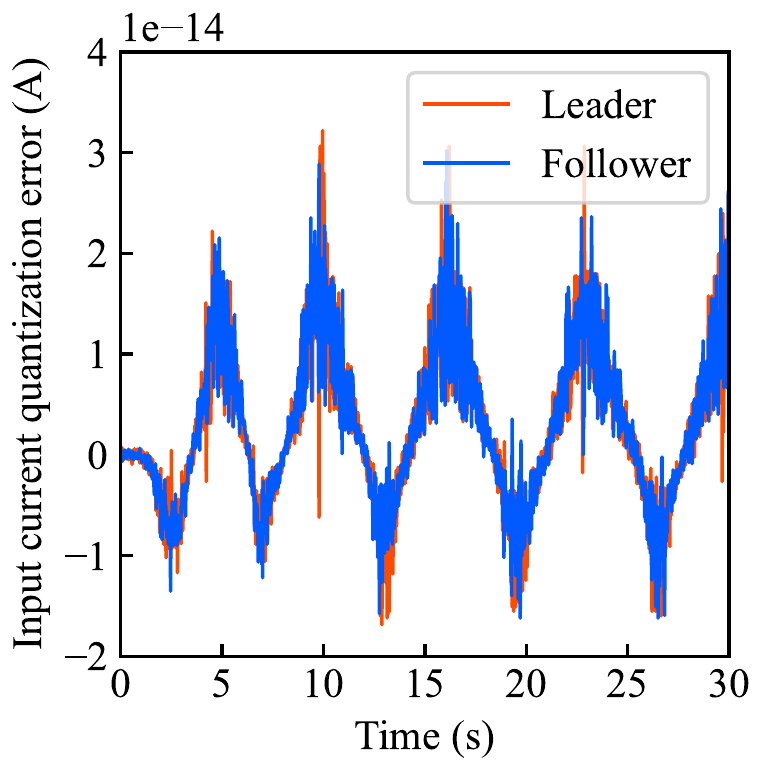}%
\label{fig:qey}}
\subfloat[]{\includegraphics[scale=.54]{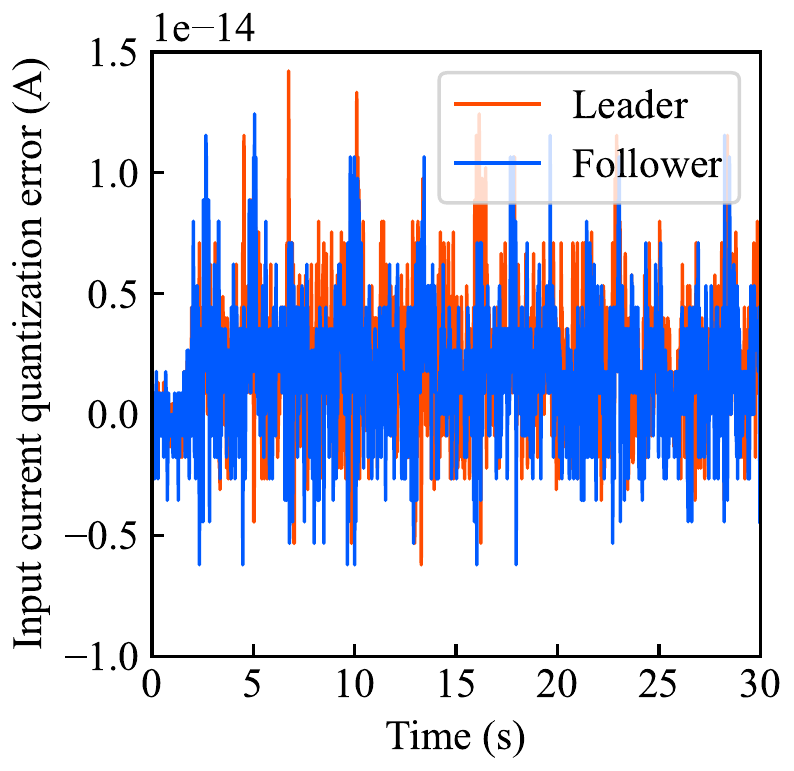}%
\label{fig:qep}}
\caption{Quantization error of input current for each motor. (a) Quantization error of input current for yaw axis motor. (b) Quantization error of input current for pitch axis motor.}
\label{fig:qe}
\end{figure}

%%%
\subsection{Effects of the controller encryption}\label{sec:qerror}
This section examines the effects of controller encryption on the real-time property and quantization error compared with unencrypted four-channel bilateral controls.
The resulting computation times and quantization errors are shown in Figs.~\ref{fig:ct} and \ref{fig:qe}, respectively.

Figs.~\ref{fig:ct}(a) and (b) show the computation times of the leader and follower when the remote control is performed with and without encryption, where the red and blue lines represent the leader and follower arms, respectively.
Comparing Figs.~\ref{fig:ct} (a) and (b), it is confirmed that the computation time in the encrypted case is on average $\SI{1.9}{ms}$ longer than that in the unencrypted case, while it is shorter than the sampling period $\SI{20}{ms}$, which implies that the developed control system ensures real-time properties.

Quantization errors are computed with respect to the control input for each motor of the leader and follower.
The quantization error, denoted as $\delta$, is defined as the difference between the output of the unencrypted controller $i$ and the output of the encrypted controller $\tilde{i}$ for the common input variables $\xi$ to the controller, that is,
\begin{align}
    \delta_k = i_k-\tilde{i}_k,\quad \forall k\in\mathbb{Z}.
\end{align}
The quantization errors for each motor are presented in Fig.~\ref{fig:qe}, where the red and blue lines represent the leader and follower arms, respectively.
Figs.~\ref{fig:qe}(a) and (b) show the quantization error of the input current on the yaw and pitch axes, respectively.
The figures show that the errors are on the order of $\SI{1e-14}{A}$ for both the yaw and pitch axes. 
The resolution of the DA board is $\SI{16}{bit}$ and the full-scale voltage is $\SI{10}{V}$, as shown in TABLE~\ref{tab:spec}.
In this case, the resolution of the voltage is ${10}/{2^{16}} \approx 1.53\times 10^{-4}\,\si{V}$, the output voltage of the DA board per current $\SI{1}{A}$ is $8/9\ $V, and the current resolution is $1.53\times 10^{-4} \times 8/9 = 1.36\times 10^{-4}\,\si{A}$.
Because $\delta$ is $10^{-10}$ smaller than the resolution of the DA board, the quantization error is negligible.
Thus, the real-time property is guaranteed by controller encryption, and the effect of quantization errors on the control input current is negligible.

%%%
\section{Conclusion\label{sec:conclusion}}
In this study, an encrypted four-channel bilateral control system with two-axis robotic arms was developed. 
A model-based four-channel bilateral control system was designed using the model parameters of the two-axis robot arm and disturbance and reaction force observers. 
The model parameters were identified through static and constant-speed tests. 
The leader and follower controllers in the bilateral control system were represented in a state-space representation, and secure implementation of the controllers was performed using a multiplicative homomorphic encryption scheme. 
The advantage of the developed control system is that it can operate using encrypted communication data and control parameters. 
Moreover, we experimentally evaluated the developed control system in terms of posture synchronization and force feedback. 
Motion-tracking experiments confirmed that the postures of the leader and the follower were synchronized. 
Obstacle contact experiments confirmed a difference in hardness between aluminum prism and sponge. 
We also confirmed that the quantization error was negligibly small and that the developed bilateral control system ran in real-time. 
Through experimental verification, the encrypted four-channel bilateral control system enables inheriting control performance from the original (unencrypted) bilateral control system. 

In the future, we will develop a bilateral control system using dynamic key encryption~\cite{Teranishi2020st} to further enhance cybersecurity.
This development expects us to protect the bilateral control system from cyber-attacks aimed at destabilizing and compromising a plant.
We will conduct a secure implementation of different bilateral control methods considering time-varying delays and losses over communication links~\cite{Anderson1989} to expand the applied situations.
Moreover, we will work on the implementation of the developed control system for industrial robotic control systems to investigate its practicability.

\bibliographystyle{IEEEtran}
\bibliography{./bib/manuscript_v5}

\begin{thebibliography}{10}
\providecommand{\url}[1]{#1}
\csname url@samestyle\endcsname
\providecommand{\newblock}{\relax}
\providecommand{\bibinfo}[2]{#2}
\providecommand{\BIBentrySTDinterwordspacing}{\spaceskip=0pt\relax}
\providecommand{\BIBentryALTinterwordstretchfactor}{4}
\providecommand{\BIBentryALTinterwordspacing}{\spaceskip=\fontdimen2\font plus
\BIBentryALTinterwordstretchfactor\fontdimen3\font minus
  \fontdimen4\font\relax}
\providecommand{\BIBforeignlanguage}[2]{{%
\expandafter\ifx\csname l@#1\endcsname\relax
\typeout{** WARNING: IEEEtran.bst: No hyphenation pattern has been}%
\typeout{** loaded for the language `#1'. Using the pattern for}%
\typeout{** the default language instead.}%
\else
\language=\csname l@#1\endcsname
\fi
#2}}
\providecommand{\BIBdecl}{\relax}
\BIBdecl

\bibitem{Clement1985}
G.~Clement, J.~Vertut, R.~Fournier, B.~Espiau, and G.~Andre, ``An overview of
  cat control in nuclear services,'' in \emph{Proc. 1985 IEEE Int. Conf. Robot.
  Autom.}, vol.~2, 1985, pp. 713--718.

\bibitem{Bejczy1987}
A.~Bejczy and Z.~Szakaly, ``Universal computer control systems (uccs) for space
  telerobots,'' in \emph{Proc. 1987 IEEE Int. Conf. Robot. Autom.}, vol.~4,
  1987, pp. 318--324.

\bibitem{Lei2011}
L.~Li, Q.~Wei, Z.~Hou, and L.~Zhao, ``Design and realization of the
  experimental platform of space robot bilateral teleoperation system,'' in
  \emph{Proc. 30th Chin. Control Conf.}, 2011, pp. 3968--3972.

\bibitem{Funda1991}
J.~Funda and R.~Paul, ``A symbolic teleoperator interface for time-delayed
  underwater robot manipulation,'' in \emph{Proc. OCEANS 91}, 1991, pp.
  1526--1533.

\bibitem{Rovetta1996}
A.~Rovetta, R.~Sala, X.~Wen, and A.~Togno, ``Remote control in telerobotic
  surgery,'' vol.~26, no.~4, pp. 438--444, 1996.

\bibitem{Jia2015}
F.~Jia, S.~Guo, and Y.~Wang, ``An interventional surgical robot system with
  force feedback,'' in \emph{Proc. 2015 IEEE Int. Conf. Robot. Biomim.}, 2015,
  pp. 632--637.

\bibitem{HOKAYEM2006}
P.~F. Hokayem and M.~W. Spong, ``Bilateral teleoperation: An historical
  survey,'' \emph{Automatica}, vol.~42, no.~12, pp. 2035--2057, 2006.

\bibitem{Niemeyer1991}
G.~Niemeyer and J.-J. Slotine, ``Stable adaptive teleoperation,'' vol.~16,
  no.~1, pp. 152--162, 1991.

\bibitem{Tachi1991}
S.~Tachi, ``Force and/or impedance control system in master-slave teleoperation
  (in japanese),'' \emph{J. Robot. Soc. Japan}, vol.~9, no.~6, pp. 766--772,
  1991.

\bibitem{Hannaford1988}
B.~Hannaford and R.~Anderson, ``Experimental and simulation studies of hard
  contact in force reflecting teleoperation,'' in \emph{Proc. IEEE Int. Conf.
  Robot. Autom.}, vol.~1, 1988, pp. 584--589.

\bibitem{Willaert2009}
B.~Willaert, B.~Corteville, D.~Reynaerts, H.~Van~Brussel, and
  E.~Vander~Poorten, ``Bounded environment passivity of the classical
  position-force teleoperation controller,'' in \emph{Proc. 2009 IEEE Int.
  Conf. Intell. Robots Syst.}, 2009, pp. 4622--4628.

\bibitem{Lawrence1993}
D.~Lawrence, ``Stability and transparency in bilateral teleoperation,'' vol.~9,
  no.~5, pp. 624--637, 1993.

\bibitem{Yokokhji1994}
Y.~Yokokohji and T.~Yoshikawa, ``{Bilateral control of master-slave
  manipulators for ideal kinesthetic coupling-formulation and experiment},''
  vol.~10, no.~5, pp. 605--620, 1994.

\bibitem{Matsumoto:2003aa}
Y.~Matsumoto, S.~Katsura, and K.~Ohnishi, ``{An analysis and design of
  bilateral control based on disturbance observer},'' in \emph{Proc. IEEE Int.
  Conf. Ind. Tech.}, vol.~2, 2003, pp. 802--807.

\bibitem{Iida2004}
W.~Iida and K.~Ohnishi, ``Reproducibility and operationality in bilateral
  teleoperation,'' in \emph{Proc. 8th IEEE Int. Workshop Adv. Motion Control},
  2004, pp. 217--222.

\bibitem{Natori2006}
K.~Natori and K.~Ohnishi, ``{A design method of communication disturbance
  observer for time delay compensation},'' in \emph{Proc. Ind. Electron.
  Conf.}, 2006, pp. 730--735.

\bibitem{Yang2020}
Y.~Yang, D.~Constantinescu, and Y.~Shi, ``Robust four-channel teleoperation
  through hybrid damping-stiffness adjustment,'' vol.~28, no.~3, pp. 920--935,
  2020.

\bibitem{Hangai2021}
S.~Hangai and T.~Nozaki, ``Haptic data prediction and extrapolation for
  communication traffic reduction of four-channel bilateral control system,''
  vol.~17, no.~4, pp. 2611--2620, 2021.

\bibitem{Natori2004}
K.~Natori, T.~Tsuji, K.~Ohnishi, A.~Hace, and K.~Jezernik, ``Robust bilateral
  control with internet communication,'' in \emph{Proc. 30th Annu. Conf. IEEE
  Ind. Electron. Soc.}, vol.~3, 2004, pp. 2321--2326.

\bibitem{Ueda2004}
J.~Ueda and T.~Yoshikawa, ``Force-reflecting bilateral teleoperation with time
  delay by signal filtering,'' vol.~20, no.~3, pp. 613--619, 2004.

\bibitem{Varkonyi2014}
T.~A. Várkonyi, I.~J. Rudas, P.~Pausits, and T.~Haidegger, ``Survey on the
  control of time delay teleoperation systems,'' in \emph{Proc. IEEE 18th Int.
  Conf. Intell. Engr. Syst.}, 2014, pp. 89--94.

\bibitem{islam2015}
S.~Islam, P.~X. Liu, A.~E. Saddik, and Y.~B. Yang, ``Bilateral control of
  teleoperation systems with time delay,'' vol.~20, no.~1, pp. 1--12, 2015.

\bibitem{Beerens2020}
R.~Beerens, D.~Heck, A.~Saccon, and H.~Nijmeijer, ``The effect of controller
  design on delayed bilateral teleoperation performance: An experimental
  comparison,'' vol.~28, no.~5, pp. 1727--1740, 2020.

\bibitem{Stuxnet2011}
T.~M. Chen and S.~Abu-Nimeh, ``Lessons from stuxnet,'' \emph{Computer},
  vol.~44, no.~4, pp. 91--93, 2011.

\bibitem{Industroyer2016}
M.~L. Robert, J.~A. Michael, and C.~Tim, ``Analysis of the cyber attack on the
  ukrainian power grid,'' \emph{Electricity Information Sharing and Analysis
  Center}, 2016.

\bibitem{Munteanu2018}
A.~Munteanu, R.~Muradore, M.~Merro, and P.~Fiorini, ``On cyber-physical attacks
  in bilateral teleoperation systems: An experimental analysis,'' in
  \emph{Proc. 2018 IEEE Ind. Cyber-Phys. Syst.}, 2018, pp. 159--166.

\bibitem{Asif2020}
M.~R. Al~Asif and R.~Khondoker, ``Cyber security threat modeling of a
  telesurgery system,'' in \emph{Proc. 2nd Int. Conf. on Sustain. Tech. Ind.
  4.0}, 2020, pp. 1--6.

\bibitem{Bonaci2015}
T.~Bonaci, J.~Herron, T.~Yusuf, J.~Yan, T.~Kohno, and H.~J. Chizeck, ``To make
  a robot secure: An experimental analysis of cyber security threats against
  teleoperated surgical robots,'' \emph{CoRR}, vol. abs/1504.04339, 2015.

\bibitem{Dong2020}
Y.~Dong, N.~Gupta, and N.~Chopra, ``False data injection attacks in bilateral
  teleoperation systems,'' vol.~28, no.~3, pp. 1168--1176, 2020.

\bibitem{kalam2016}
A.~{Abou El Kalam}, A.~Ferreira, and F.~Kratz, ``{Bilateral Teleoperation
  System Using QoS and Secure Communication Networks for Telemedicine
  Applications},'' vol.~10, no.~2, pp. 709--720, 2016.

\bibitem{Tozal2011}
M.~E. Tozal, Y.~Wang, E.~Al-Shaer, K.~Sarac, B.~Thuraisingham, and B.-T. Chu,
  ``On secure and resilient telesurgery communications over unreliable
  networks,'' in \emph{Proc. 2011 IEEE Conf. Comp. Comm. Workshops}, 2011, pp.
  714--719.

\bibitem{Jiang2020}
Z.~Jiang, Y.~Zhang, and X.~Cai, ``Adaptive control of bilateral teleoperation
  system under denial of service attacks,'' in \emph{Proc. 39th Chin. Control
  Conf.}, 2020, pp. 4566--4571.

\bibitem{Kogiso2015}
K.~Kogiso and T.~Fujita, ``{Cyber-security enhancement of networked control
  systems using homomorphic encryption},'' in \emph{Proc. IEEE Conf. Decis.
  Control.}, 2015, pp. 6836--6843.

\bibitem{FAROKHI2016163}
F.~Farokhi, I.~Shames, and N.~Batterham, ``{Secure and Private Cloud-Based
  Control Using Semi-Homomorphic Encryption},'' \emph{IFAC-PapersOnLine},
  vol.~49, no.~22, pp. 163--168, 2016.

\bibitem{Kim2016175}
J.~Kim, C.~Lee, H.~Shim, J.~H. Cheon, A.~Kim, M.~Kim, and Y.~Song,
  ``{Encrypting Controller using Fully Homomorphic Encryption for Security of
  Cyber-Physical Systems},'' \emph{IFAC-PapersOnLine}, vol.~49, no.~22, pp.
  175--180, 2016.

\bibitem{schlze2021en}
M.~{Schulze Darup}, A.~B. Alexandru, D.~E. Quevedo, and G.~J. Pappas,
  ``{Encrypted Control for Networked Systems: An Illustrative Introduction and
  Current Challenges},'' vol.~41, no.~3, pp. 58--78, 2021.

\bibitem{Andreea2017}
A.~B. Alexandru, K.~Gatsis, and G.~J. Pappas, ``Privacy preserving cloud-based
  quadratic optimization,'' in \emph{Proc. 55th Annu. Allerton Conf. Commun.
  Control Comput.}, 2017, pp. 1168--1175.

\bibitem{shono2022}
S.~Naoto, M.~Tetsuro, T.~Kaoru, K.~Takahiro, K.~Toshihiro, K.~Kiminao, and
  K.~Kenji, ``Implementation of encrypted control of pneumatic bilateral
  control system using wave variables,'' in \emph{Proc. 27th Int. Symp. Artif.
  Life Robot.}, 2022.

\bibitem{KOSIERADZKI2022se}
S.~Kosieradzki, X.~Zhao, H.~Kawase, Y.~Qiu, K.~Kogiso, and J.~Ueda, ``Secure
  teleoperation control using somewhat homomorphic encryption,''
  \emph{IFAC-PapersOnLine}, vol.~55, no.~37, pp. 593--600, 2022.

\bibitem{Zhu1995}
M.~Zhu and S.~Salcudean, ``Achieving transparency for teleoperator systems
  under position and rate control,'' in \emph{Proc. IEEE Int. Conf. Intell.
  Robot. Syst.}, vol.~2, 1995, pp. 7--12 vol.2.

\bibitem{Teranishi2020en}
K.~Teranishi, K.~Kogiso, and J.~Ueda, ``{Encrypted feedback linearization and
  motion control for manipulator with somewhat homomorphic encryption},'' in
  \emph{Proc. IEEE/ASME Int. Conf. Adv. Intell. Mechatron.}, 2020, pp.
  613--618.

\bibitem{ARCS}
\BIBentryALTinterwordspacing
Y.~Yokokura, ``Side warehouse of laboratory.'' accessed: 2022-11-05. [Online].
  Available: \url{http:// www.sidewarehouse.net/arcs51/index.html}
\BIBentrySTDinterwordspacing

\bibitem{Murakami:1993aa}
T.~Murakami, F.~Yu, and K.~Ohnishi, ``{Torque sensorless control in
  multidegree-of-freedom manipulator},'' vol.~40, no.~2, pp. 259--265, apr
  1993.

\bibitem{Ohnishi1996mc}
K.~Ohnishi, M.~Shibata, and T.~Murakami, ``{Motion control for advanced
  mechatronics},'' vol.~1, no.~1, pp. 56--67, 1996.

\bibitem{gopinath1971}
B.~Gopinath, ``{On the control of linear multiple input-output systems},''
  \emph{Bell Syst. Tech. J.}, vol.~50, no.~3, pp. 1063--1081, 1971.

\bibitem{elgamal1985}
T.~Elgamal, ``{A public key cryptosystem and a signature scheme based on
  discrete logarithms},'' vol.~31, no.~4, pp. 469--472, 1985.

\bibitem{Teranishi2020st}
K.~Teranishi, N.~Shimada, and K.~Kogiso, ``Stability-guaranteed dynamic elgamal
  cryptosystem for encrypted control systems,'' \emph{IET Control Theory \&
  Applications}, vol.~14, no.~16, pp. 2242--2252, 2020.

\bibitem{Anderson1989}
R.~Anderson and M.~Spong, ``Bilateral control of teleoperators with time
  delay,'' vol.~34, no.~5, pp. 494--501, 1989.

\end{thebibliography}

\begin{IEEEbiography}[{\includegraphics[width=1in,height=1.25in,clip,keepaspectratio]{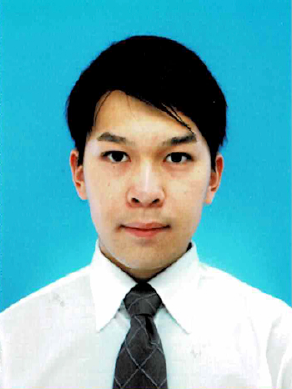}}]{Haruki Takanashi} (Student Member, {IEEE}) received the B.E. degree from The University of Electro-Communications, Tokyo, Japan.
He is currently an M.S. student at The University of Electro-Communications, Tokyo, Japan. 

His research interests include teleoperation and encrypted control.
\end{IEEEbiography}

\begin{IEEEbiography}[{\includegraphics[width=1in,height=1.25in,clip,keepaspectratio]{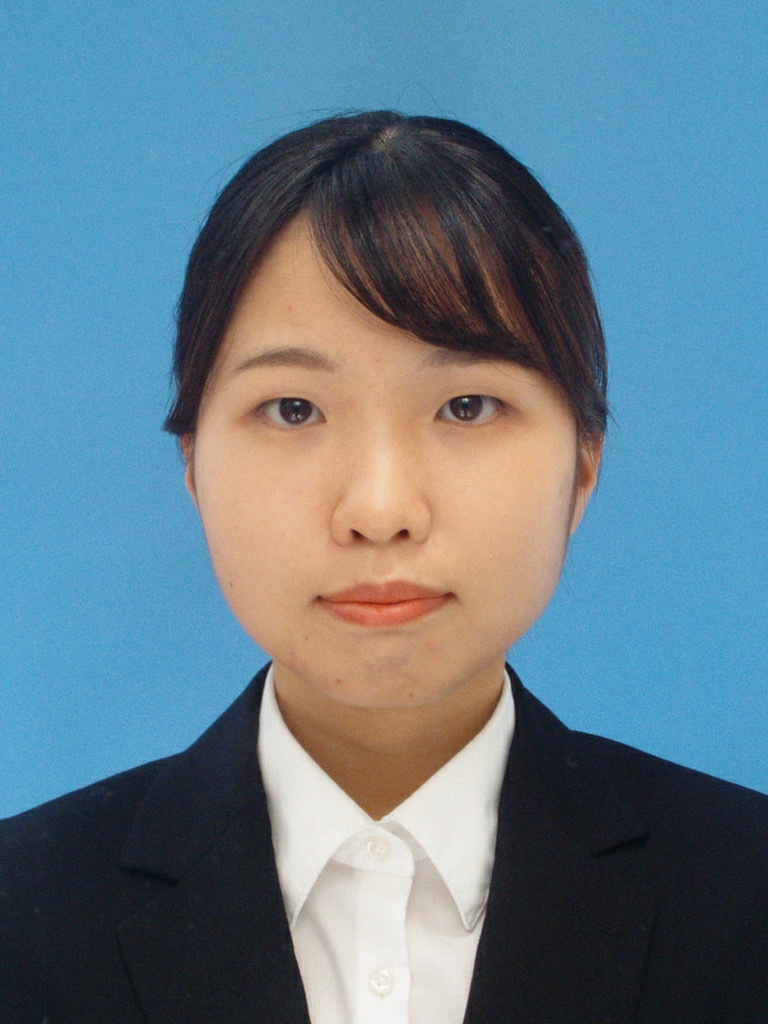}}]{Akane Kosugi} (Student Member, {IEEE}) received the B.E. degree from The University of Electro-Communications, Tokyo, Japan, in 2022. 
She is currently a graduate student in a master's course at The University of Electro-Communications. 

Her research interests include cyber-security of control systems.
\end{IEEEbiography}

\begin{IEEEbiography}[{\includegraphics[width=1in,height=1.25in,clip,keepaspectratio]{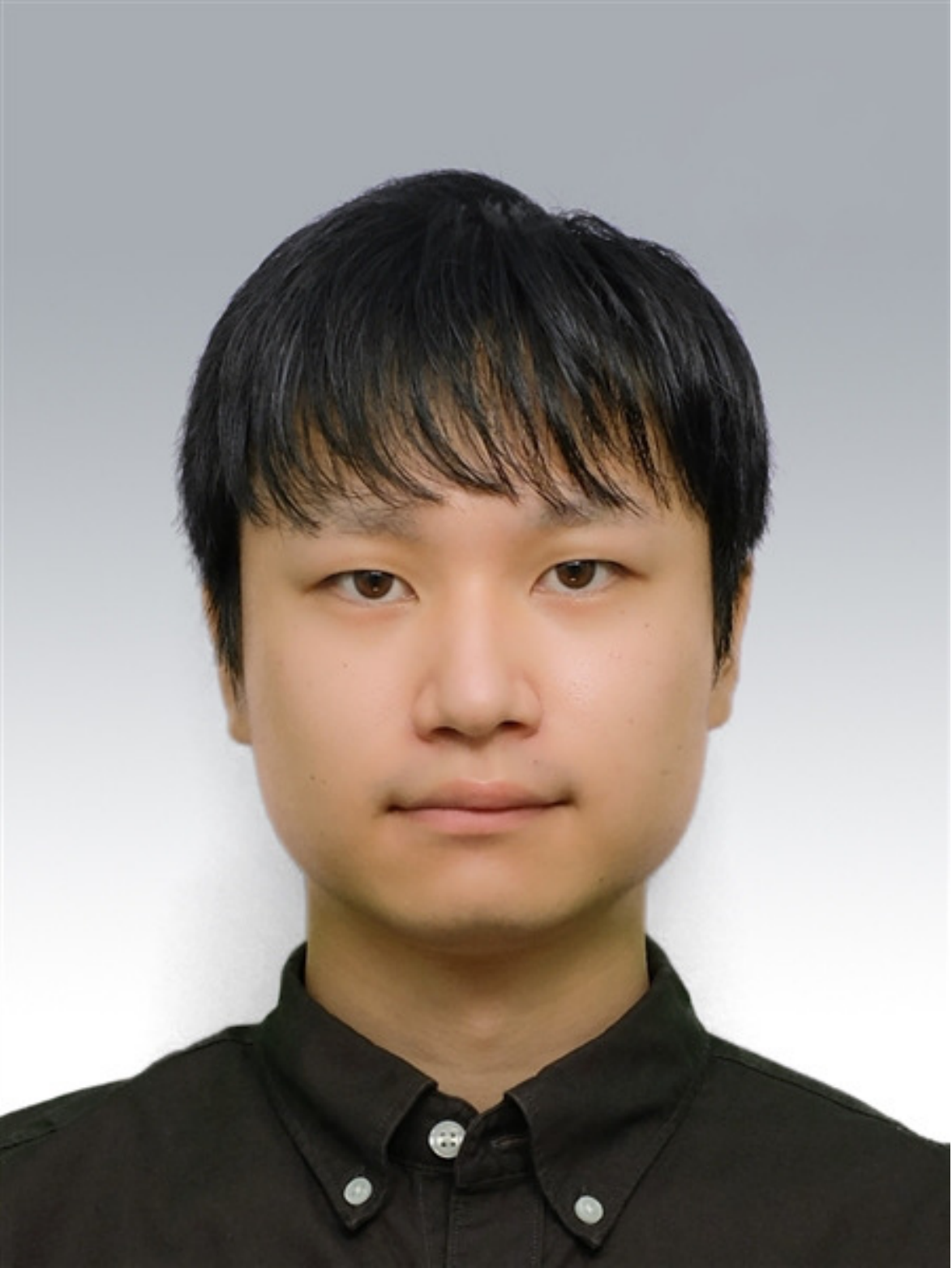}}]
    {Kaoru Teranishi} (Student Member, {IEEE}) received the B.S. degree in electromechanical engineering from National Institute of Technology, Ishikawa College, Ishikawa, Japan, in 2019.
    He also obtained the M.S. degree in Mechanical and Intelligent Systems Engineering from The University of Electro-Communications, Tokyo, Japan, in 2021.
    He is currently a Ph.D. student at The University of Electro-Communications.
    From October 2019 to September 2020, he was a visiting scholar of the Georgia Institute of Technology, GA, USA.
    Since April 2021, he has been a Research Fellow of Japan Society for the Promotion of Science.
    His research interests include control theory and cryptography for cyber-security of control systems.
\end{IEEEbiography}

\begin{IEEEbiography}[{\includegraphics[width=1in,height=1.25in,clip,keepaspectratio]{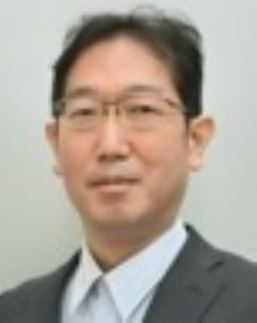}}]{Toru Mizuya} received the B.E. and M.E. degrees in Mechanical Engineering from the University of Tokyo, Tokyo, Japan, in 1996 and 1998, respectively. Currently, he is with the Local Independent Administrative Agency Kanagawa Institute of Industrial Science and Technology and engages in research and development related to the application of 5G wireless communication.  

His research interests include software engineering, system engineering, information models, programming languages, and so on. 
\end{IEEEbiography}

\begin{IEEEbiography}[{\includegraphics[width=1in,height=1.25in,clip,keepaspectratio]{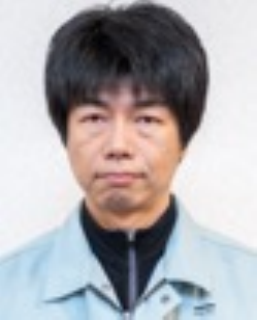}}]{Kenichi Abe} received the M.E. degree from Musashi Institute of Technology, Tokyo, Japan, in 1991.
Currently, he is a researcher in the Information and Production Engineering Department, Local Independent Administrative Agency Kanagawa Institute of Industrial Science and Technology, Ebina, Japan. 

His research interests include mechanical engineering and instrumentation engineering.
\end{IEEEbiography}

\begin{IEEEbiography}[{\includegraphics[width=1in,height=1.25in,clip,keepaspectratio]{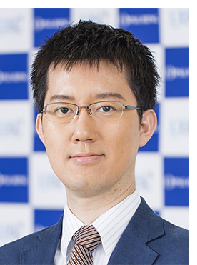}}]{Kiminao Kogiso} (Member, {IEEE}) received the B.S., M.S., and Ph.D. degrees in mechanical engineering from Osaka University, Japan in 1999, 2001, and 2004, respectively.

He was a postdoctoral researcher in the 21st Century COE Program in 2004 and became an Assistant Professor in the Department of Information Systems, Nara Institute of Science and Technology, Nara, Japan, in 2005.
Since March 2014, he has been an Associate Professor in the Department of Mechanical and Intelligent Systems Engineering, The University of Electro-Communications, Tokyo, Japan.
From November 2010 to December 2011, he was a visiting scholar at Georgia Institute of Technology, GA, USA.

His research interests include constrained control, control of decision makers, cyber-security of control systems, and their applications.
\end{IEEEbiography}

\vfill

\end{document}